%% file: main.tex
\documentclass[10pt,conference]{IEEEtran}
\IEEEoverridecommandlockouts
\usepackage{cite}
\usepackage{amsmath,amssymb,amsfonts}
\usepackage{algorithmic}
\usepackage{graphicx}
\usepackage{textcomp}
\usepackage{xcolor}
\usepackage[most]{tcolorbox}
\usepackage{booktabs}
\usepackage{threeparttable}
\usepackage{enumitem}
\usepackage{url}
\usepackage{hyperref}
\def\BibTeX{{\rm B\kern-.05em{\sc i\kern-.025em b}\kern-.08em
    T\kern-.1667em\lower.7ex\hbox{E}\kern-.125emX}}
\begin{document}

\title{Hierarchical Knowledge Injection for Improving LLM-based Program Repair}

\author{
\IEEEauthorblockN{Ramtin Ehsani}
\IEEEauthorblockA{
\textit{Drexel University} \\
Philadelphia, PA, USA \\
ramtin.ehsani@drexel.edu}
\and
\IEEEauthorblockN{Esteban Parra}
\IEEEauthorblockA{
\textit{Belmont University} \\
Nashville, TN, USA \\
esteban.parrarodriguez@belmont.edu}
\and
\IEEEauthorblockN{Sonia Haiduc}
\IEEEauthorblockA{
\textit{Florida State University} \\
Tallahassee, FL, USA \\
shaiduc@fsu.edu}
\and
\IEEEauthorblockN{Preetha Chatterjee}
\IEEEauthorblockA{
\textit{Drexel University} \\
Philadelphia, PA, USA \\
preetha.chatterjee@drexel.edu}
}

\maketitle

\begin{abstract}
%Prompting Large Language Models (LLMs) with bug-related context, such as error messages and stack traces, has been shown to improve automated program repair (APR). However, many bugs remain unresolved even with this information, suggesting a gap in how effectively LLMs understand and fix complex issues. Focusing solely on the immediate surroundings of the buggy code often fails to provide the full picture. In real-world development, resolving such bugs frequently requires a broader understanding of the repository and project, including structural dependencies, version history, documentation, and prior fixes.
Prompting LLMs with bug-related context (e.g., error messages, stack traces) improves automated program repair, but many bugs still remain unresolved. 
%This highlights a gap in LLMs' ability to handle complex issues. 
In real-world projects, developers often rely on broader repository and project-level context beyond the local code to resolve such bugs. In this paper, we investigate how automatically extracting and providing such knowledge can improve LLM-based program repair. We propose a layered knowledge injection framework that incrementally augments LLMs with structured context. It starts with the \textit{Bug Knowledge Layer}, which includes information such as the buggy function and failing tests; expands to the \textit{Repository Knowledge Layer}, which adds structural dependencies, related files, and commit history; and finally injects the \textit{Project Knowledge Layer}, which incorporates relevant details from documentation and previously fixed bugs. 
We evaluate this framework on a dataset of 314 bugs from BugsInPy using two LLMs (Llama 3.3 and GPT-4o-mini), and analyze fix rates across six bug types. %such as \textit{Network} and \textit{GUI}, at each layer of contextual injection.
By progressively injecting knowledge across layers, our approach achieves a fix rate of 79\% (250/314) using Llama 3.3, a significant improvement of 23\% over previous work. All bug types show improvement with the addition of repository-level context, while only a subset benefit further from project-level knowledge, highlighting that different bug types require different levels of contextual information for effective repair.
We also analyze the remaining unresolved bugs and find that more complex and structurally isolated bugs, such as \textit{Program Anomaly} and \textit{GUI} bugs, remain difficult even after injecting all available information. %Our findings show the importance of structured, layered context and suggest the need for interactive and adaptive strategies in future APR systems.
Our results show that layered context injection improves program repair and suggest the need for interactive and adaptive APR systems. %offer suggestions for developing adaptive, interactive APR systems.
\end{abstract}


\input{01_intro}
\input{02_motivation}

\input{03_methodology}
\input{04_results}
\input{05_threats}
\input{06_background}
\input{07_conclusions}

\bibliographystyle{IEEEtran}
\bibliography{ref}

\end{document}

%% file: 01_intro.tex
\section{Introduction}
Large Language Models (LLMs) have opened new possibilities in Automated Program Repair (APR), offering the ability to generate patches with minimal supervision. Despite this promise, LLMs often struggle to grasp the deeper context behind bugs, leading to unreliable or non-applicable fixes~\cite{10.1145/3597503.3639086, 10.1109/ASE56229.2023.00047, 10.1145/3715004}. A key limitation is that many generated patches lack awareness of project-level constraints, dependencies, or intent~\cite{10.1145/3395363.3397369, 10.1145/3180155.3180233}. Recent findings show that the performance of LLMs drops by more than 50\% when bug fixes require information beyond the buggy function~\cite{chen2025studyingunderstandingeffectivenessfailures}. Building effective LLM-based tools for APR requires a better understanding of when and why these models succeed or fail, an area that is still underexplored.

%generating patches that are accurate in context and align with the overall structure of the codebase~\cite{10.1145/3395363.3397369, 10.1145/3180155.3180233}.
% Due to the context window limitations of LLMs, it is impractical to provide an entire project’s codebase as input.
% often lacks focus—being either too verbose or overly generic

Research has shown that incorporating additional relevant information in prompts can enhance the performance of LLMs in program repair. 
Zhao et al.\cite{10.1145/3691620.3695537} extract design rationales from issue threads, but the approach relies heavily on developer discussions, which are often noisy or inconsistent~\cite{10.1145/3691620.3695019}.
%Zhao et al.~\cite{10.1145/3691620.3695537} leveraged DRMiner~\cite{10.1145/3691620.3695019} to extract design rationales from developer discussions in issue threads. %These rationales capture high-level reasoning on how to approach and resolve coding problems, leading to improved program repair performance. 
%While promising, this method relies on the quality of the discussions, which are often noisy or inconsistent~\cite{10.1145/3691620.3695019}. %, and ultimately requires developer input.
Parasaram et al.~\cite{parasaram2024factselectionproblemllmbased} %extended this idea with fact selection techniques aimed at improving LLM performance in APR. They demonstrated that 
observed improved LLM performance in APR by incorporating \textit{bug-related facts}, such as error messages, stack traces, and buggy code snippets. 
%HAFix uses commit history to examine how past commits in the repository could improve LLM-based program repair~\cite{shi2025hafix}. 
HAFix explored commit history to provide historical context for buggy functions, showing improvements on a small dataset~\cite{shi2025hafixhistoryaugmentedlargelanguage}.
%By comparing changes in commits made to buggy functions over time, HAFix provides LLMs with insights into what went wrong in the code. Although on a very small dataset of 51 bugs, LLMs demonstrated better performance when provided with the history of project commits. 
While these approaches highlight the value of adding relevant context, most are limited to the immediate surroundings of the buggy code. %There is a missed opportunity to leverage other important signals, such as structural dependencies and documentation, found at the \textit{repository} and \textit{project} levels. 
%Even those that go beyond lack integration of structural dependencies or project-specific knowledge. 
%While some approaches, like HAFix, extend beyond local code by incorporating historical commit data, they still do not account for the entire structure or semantics of the repository and lack integration of other rich sources of project-specific knowledge. %This localized view can limit an LLM's ability to resolve complex bugs that require a broader understanding of the system.
% their scope remains narrowly confined to the immediate vicinity of the buggy code.
%This localized focus ignores the broader project context, which may contain essential information needed for resolving complex software bugs.
It is known that bugs often span multiple files and components~\cite{1510173}. Thus, including a broader context, such as related files, dependencies, and documentation, could produce more accurate patches. However, 
Parasaram et al.~\cite{parasaram2024factselectionproblemllmbased} showed that there is no universal set of information that works across all scenarios and bugs. 
Their tool uses a Random Forest model to select types of contextual information to include in the prompt based on features such as prompt length, repository ID, and code complexity. However, their model lacks interpretability and ignores bug type. Different bug types require different contextual cues to be solved~\cite{CATOLINO2019165}, and treating them uniformly could lead to producing ineffective or suboptimal repairs.

%Research in large-scale software systems indicates that bugs are often interconnected across multiple files and components within the project~\cite{1510173}. This inter-connectivity suggests that providing a more comprehensive view of the project's context, beyond just the buggy code, could significantly enhance an LLM’s ability to generate fixes that align with the project's overall structure and requirements. Expanding the scope to include related files, dependencies, and relevant documentation might enable LLMs to produce more accurate and contextually appropriate fixes. In addition, there is no one-size-fits-all approach to context in program repair. Parasaram et al.~\cite{parasaram2024factselectionproblemllmbased} showed that a universally applicable subset of facts does not exist. Their use of a Random Forest model in a tool called MANIPLE, which selects facts based on features such as prompt length, repository ID, and cyclomatic complexity of the buggy code, is a step forward. However, this approach suffers from two key limitations: First, it relies on a black-box model, making it difficult to interpret which features are truly impactful for fact selection. Second, it fails to account for the type of bug causing the issue. Research has consistently shown that not all bugs are created equal; each bug type has unique characteristics that affect how it should be addressed~\cite{CATOLINO2019165}. Ignoring this diversity in bug types could lead to suboptimal solutions, as the context required to resolve different kinds of bugs will inevitably vary.

In this paper, we investigate how the injection of \textit{repository} and \textit{project}-specific contextual information on top of information surrounding buggy code affects the ability of LLMs to repair different types of bugs. We use a subset of the BugsInPy dataset~\cite{10.1145/3368089.3417943}, consisting of 314 bug–patch pairs from Python open-source repositories on GitHub. Using an existing taxonomy~\cite{CATOLINO2019165}, we manually categorized each bug into one of six types, including \textit{Program Anomaly}, \textit{GUI}, \textit{Network}, etc.
We organize the contextual information into three layers: \textit{Bug Knowledge}, \textit{Repository Knowledge}, and \textit{Project Knowledge}. By progressively injecting knowledge from each layer, we aim to understand which kinds of context can help fix specific types of bugs. We evaluate our approach using two LLMs (Llama 3.3 and GPT-4o-mini), demonstrating that improvements from layered knowledge injection are consistent across models with different architectures and parameter sizes. Additionally, we perform a detailed error analysis to uncover why LLMs struggle with specific types of bugs, even after providing contextual information. Our research questions are:

\noindent
\textbf{RQ1: To what extent can LLMs resolve different types of bugs using only immediate bug knowledge?} 
%We classify each bug using Catolino et al.’s taxonomy~\cite{CATOLINO2019165}, and 
We inject bug-related context, adopted from Parasaram et al.~\cite{parasaram2024factselectionproblemllmbased}, as the \textit{Bug Knowledge Layer} to assess LLMs' performance in bug repair for different types of bugs. 
%With only immediate bug context, 
Our best-performing model fixes 207/314 bugs (65\%), surpassing the previously reported fix rate of 56\%~\cite{parasaram2024factselectionproblemllmbased}. 
This improved performance could be attributed to structured in-context knowledge injection as well as using newer versions of LLMs (Llama3.3 vs. Llama3). However, our fix rate leaves room for improvement, indicating that 
%This demonstrates that structured in-context knowledge injection, even at the bug level, can significantly improve performance. However, 
local context alone is often insufficient, especially for complex bug types like \textit{Program Anomaly} and \textit{Network}, which frequently remain unresolved in this layer.

\noindent
\textbf{RQ2: How does injecting repository-level knowledge enhance LLM-based repair of different bug types?}
We re-patch all unresolved bugs from RQ1 after adding additional context from the \textit{Repository Knowledge Layer}, including co-occurring files, structural dependencies, and commit history. This leads to a significant improvement of 9\% for our best model, fixing 28 more bugs and reaching a 74\% fix rate (235/314). All bug types benefit from this layer, showing that repository-level context is crucial for resolving bugs that local information alone cannot address.

\noindent
\textbf{RQ3: How does injecting project-level knowledge further enhance LLM-based repair of different bug types?} 
We further augment the information available for unresolved bugs from RQ2 with additional context from the \textit{Project Knowledge Layer}, including documentation and previously resolved issues. This results in our best model fixing 15 more bugs, reaching a final fix rate of 79\%  overall (250/314), a significant improvement of 23\% over previous work~\cite{parasaram2024factselectionproblemllmbased}. Improvements in this layer are limited to \textit{Program Anomaly}, \textit{GUI}, and \textit{Network} bugs, suggesting that project knowledge offers benefits for specific bug types.

\noindent
%\textbf{RQ4: Which types of bugs remain challenging for LLMs to solve, even after all layered contextual knowledge is injected?}
\textbf{RQ4: Which bug types remain challenging for LLMs, even after injecting all layers of contextual knowledge?}
We analyze the remaining unresolved bugs after all context layers are applied. We found that these bugs often require reasoning about user-facing behavior or implicit runtime conditions and feedback. These bugs also exhibit higher complexity, suggesting that current LLMs still struggle with structurally isolated or complex repair tasks.

The key contributions in this paper are:
\begin{itemize}[leftmargin=*]
    \item A layered knowledge injection framework for LLM-based APR, structuring contextual information into three hierarchical levels: \textit{Bug}, \textit{Repository}, and \textit{Project} knowledge.
    \item Evaluation against two baselines: (1) a prior fine-grained fact selection approach by Parasaram et al.~\cite{parasaram2024factselectionproblemllmbased}, and (2) an ``all-at-once" baseline where all contextual information is injected simultaneously.
    \item A bug-type-specific analysis showing how different context layers impact the repair of different types of bugs.
    \item An error analysis (e.g., bug types, complexity) of unresolved bugs, offering insights into the limitations of current LLMs and outlining concrete directions for future work.
    \item Replication package\footnote{\textbf{https://github.com/SOAR-Lab/llm-apr-knowledge-injection}} with our bug-type annotations, prompt designs, and evaluation pipeline to support reproducibility and future work on context-aware APR with LLMs.
\end{itemize}

%% file: 02_motivation.tex
\begin{figure}[th!]
    \centering
    \includegraphics[width=0.48\textwidth]{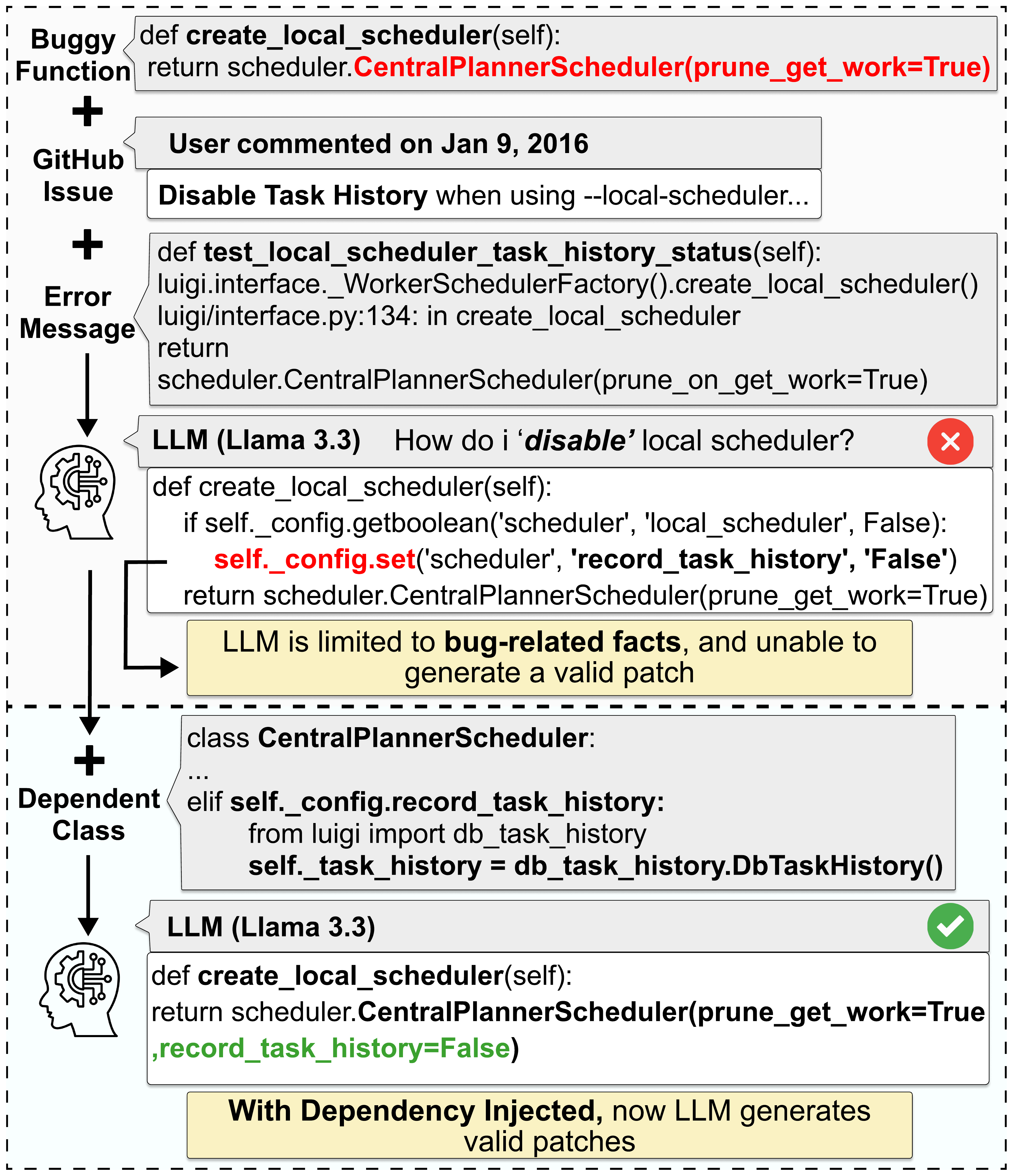}
    \vspace{-0.3cm}
    \caption{Comparison of Patch Generation with Different Information}
    \label{fig:motivate_combined}
    \vspace{-0.5cm}
\end{figure}

\section{Motivating Example}
Prior work has shown that incorporating contextual information (e.g., stack traces) in prompts enhances the bug-fixing performance of LLMs~\cite{10.1145/3180155.3180233}. Building on this work, Parasaram et al.~\cite{parasaram2024factselectionproblemllmbased} introduced a set of \textit{bug-related facts} to provide additional context for LLMs. These facts include:
\textit{Buggy Function}, \textit{Failing Tests}, \textit{Error Information}, \textit{Runtime Information}, \textit{Angelic Forest}, \textit{Buggy Class Declaration}, \textit{Method Signatures}, and \textit{GitHub Description}. Including such facts significantly improved LLM-generated fixes compared to prompting with no additional context.
However, their approach achieved a maximum fix rate of 56\% (177/314 bugs in  BugsInPy). This raises the question: What about the remaining 137 bugs? What prevented the LLM from fixing them despite the addition of varied \textit{bug-related contextual information}?

\begin{figure*}[ht!]
    \centering
    \includegraphics[width=0.95\textwidth]{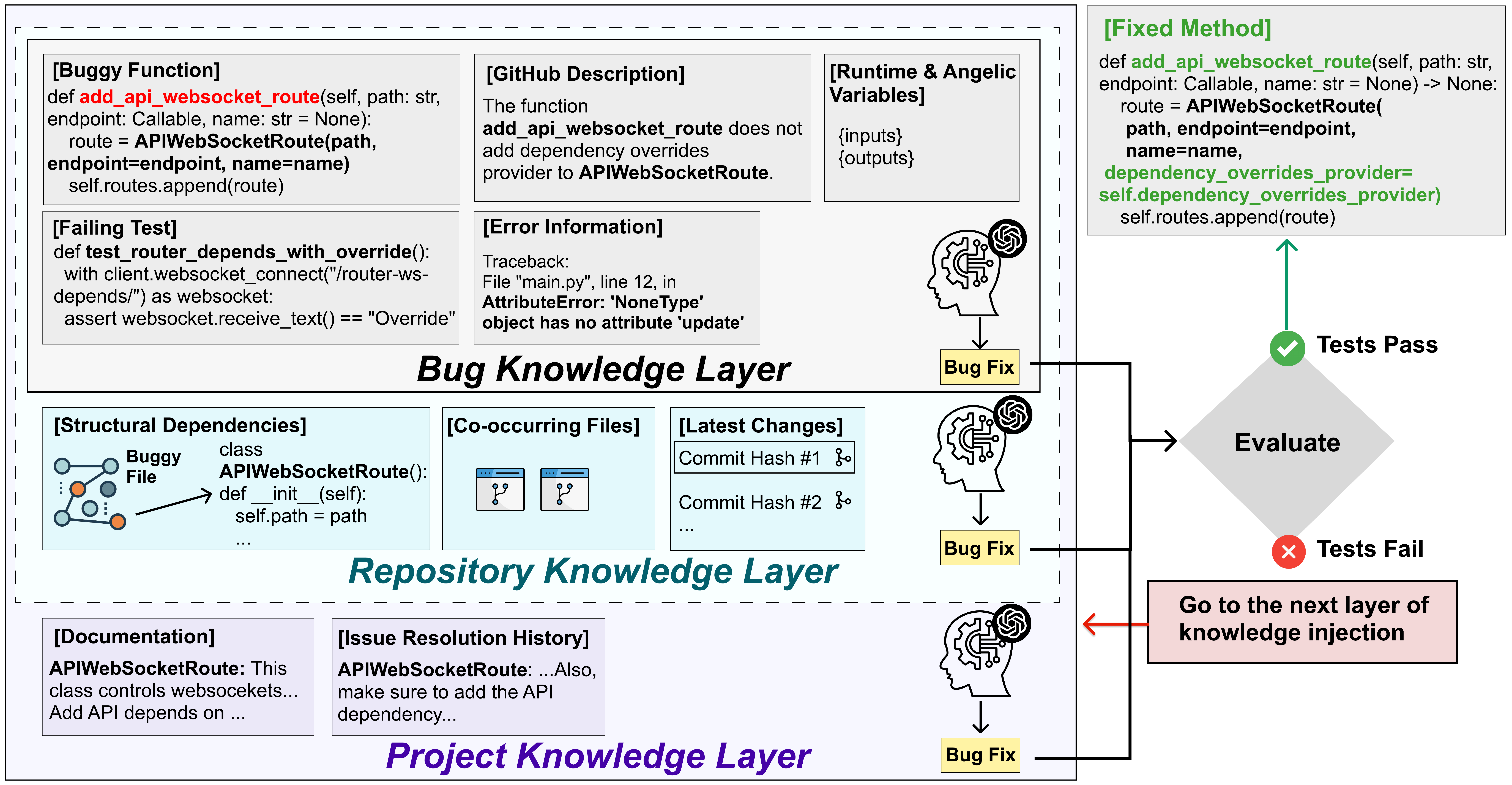}
    \vspace{-0.2cm}
    \caption{Layered Knowledge Injection for Automated Program Repair with LLMs}
    \label{fig:tool}
\vspace{-0.4cm}
\end{figure*}

We hypothesize that the key missing piece is the broader \textit{repository and project-level understanding}. 
In previous work, contextual signals are limited to the buggy function and its immediate surroundings, thus overlooking rich sources of information embedded in code structure, version history, and documentation. These are resources that developers rely on when fixing real-world bugs~\cite{10.1109/ICSE.2009.5070530, 6227188, 4016573}. 

Consider the example in Figure~\ref{fig:motivate_combined}, a bug in the \textit{luigi} project (\href{https://github.com/spotify/luigi/issues/1496}{\textcolor{blue}{link}}).
This bug, located in the \textit{create\_local\_scheduler} function, can be fixed by disabling task history. While the fix may appear trivial to a developer familiar with the codebase, Llama 3.3 fails to generate the correct patch when prompted only with the \textit{bug-related facts}. It cannot determine which class attribute to modify and instead hallucinates the existence of a variable \textit{self.\_config}. However, once dependent class information \textit{CentralPlannerScheduler} (extracted from related files) is injected into the prompt, the model produces the correct fix by disabling the \textit{record\_task\_history} attribute. This shows how structural context improves the model’s understanding.
More importantly, this example showcases that bugs do not exist in isolation, and solving many of them requires knowledge that extends beyond the local scope, spanning related files, structural dependencies, and even discussions buried in documentation or past issue threads. As we show in later sections, this is particularly true for complex bug types such as \textit{GUI}, \textit{Program Anomaly}, and \textit{Configuration} bugs.

Overall, the landscape of LLM-guided APR is fast evolving. 
%has changed since Parasaram et al.’s study. Recent advances in LLMs now support context windows of up to \textit{128k} tokens, and both small (e.g., GPT-4o-mini) and bigger models (e.g., Llama 3.3) have become significantly more capable at handling long and complex inputs. These advancements shift the core challenge in APR. 
With the advent of large-context LLMs (e.g., 128k token window in GPT-4o-mini), the challenge has shifted from fitting a small subset of facts into limited prompts to \textit{understanding what information is useful and how to strategically provide it as model input to guide effective repair.}
%The problem is no longer about selecting a small subset of facts that fit into a tight context window since newer LLMs offer expanded prompt windows (e.g., 128k tokens for GPT-40-mini). Instead, it is about how to strategically \textit{inject} relevant information to guide the model toward effective repair. Furthermore, 
Recent research also suggests that \textit{knowledge injection} (providing relevant in-context learning at inference time) can be more effective and generalizable than fine-tuning~\cite{ovadia-etal-2024-fine}.
However, providing too much information or unfiltered context may not help and can even be detrimental in APR~\cite{parasaram2024factselectionproblemllmbased}. Therefore, we need structured ways to select, automatically extract, and prioritize the most relevant information from vast and noisy software ecosystems. %Our work is motivated by the observation that real-world bug repair %often requires far more than a snapshot of buggy code. It requires understanding the ecosystem the code lives in, and we aim to automatically extract this knowledge into the prompt.

%% file: 03_methodology.tex
\section{Methodology}
%Our research addresses this gap by developing %an \textit{In-context Learning (ICL)} framework. %ICL offers a more flexible and scalable alternative to few-shot learning by extracting, organizing, and feeding into LLMs structured domain-specific signals. 
We propose a \textit{layered knowledge injection pipeline} for LLM-based bug repair. Figure~\ref{fig:tool} presents an overview of our approach. We incrementally inject contextual information through three distinct layers. 
(1) \textbf{Bug Knowledge Layer}: We begin by injecting local bug-related information (e.g., buggy function, error message, GitHub issue description). This lightweight context is often sufficient for fixing straightforward bugs. 
(2) \textbf{Repository Knowledge Layer}: For bugs that remain unresolved after layer (1), we proceed to inject repository-level context, including related files (co-commits), structural dependencies, and recent commit history. This helps the model understand a broader context beyond localized code snippets and error information. %structure and logic of the codebase. 
(3) \textbf{Project Knowledge Layer}: For the bugs still unresolved even after layer (2), we further inject knowledge extracted from documentation and previously resolved bugs. These sources provide project-level insights, such as intended system behavior, fix patterns, and edge cases. %not available in code alone.

\textit{This layered approach offers several advantages.} First, it allows simpler bugs to be fixed with minimal input, conserving tokens and computation. Second, it scales context progressively, injecting more information only when necessary. Third, it enables analysis of which bug types benefit from specific contextual signals, offering insights into LLM capabilities across different bug categories and levels of complexity. 

\begin{table*}[ht!]
\caption{Descriptions, Examples, and Frequency of Bug Types  in our Dataset}
\vspace{-0.2cm}
\label{table:dataset}
\resizebox{\textwidth}{!}{%
\begin{tabular}{|l|l|l|c|}
\hline
\textbf{Bug} & \textbf{Description} & \textbf{Example} & \textbf{Count} \\ \hline
\textbf{Program Anomaly} & \begin{tabular}[c]{@{}l@{}}Concerned with specific circumstances such as exceptions, problems with return values, \\ and unexpected crashes due to issues in the logic (rather than, e.g., the interface) of the program\end{tabular} & \begin{tabular}[c]{@{}l@{}}``Program terminates prematurely before\\ all execution events are loaded in the model"\end{tabular} & 187 \\ \hline
\textbf{Network} & \begin{tabular}[c]{@{}l@{}}Bugs having as root cause connection or server issues, due to network problems, unexpected\\ server shutdowns, or communication protocols that are not properly used within the source code\end{tabular} & \begin{tabular}[c]{@{}l@{}}``During a recent reorganization of code a\\ couple of weeks ago, SSL recording no\\ longer works"\end{tabular} & 47 \\ \hline
\textbf{Configuration} & \begin{tabular}[c]{@{}l@{}}Bugs concerned with building configuration files. Most of them are related to problems caused\\ by (i) external libraries that should be updated or fixed and (ii) wrong directories, file paths,\\ or patterns in XML or manifest artifacts\end{tabular} & \begin{tabular}[c]{@{}l@{}}``JEE5 Web model does not update on\\ changes in web.xml"\end{tabular} & 34 \\ \hline
\textbf{GUI-Related} & \begin{tabular}[c]{@{}l@{}}Bugs occurring in the Graphical User Interface of a software. It includes issues referring to \\ (i) stylistic errors, i.e., screen layouts, element colors and padding, text box appearance, and buttons, \\ as well as (ii) unexpected failures appearing to the  users in the form of unusual error messages\end{tabular} & \begin{tabular}[c]{@{}l@{}}``Text when typing in input box is not\\ viewable."\end{tabular} & 28 \\ \hline
\textbf{Performance} & \begin{tabular}[c]{@{}l@{}}Bugs reporting performance issues like memory overuse, energy leaks, and methods causing endless \\ loops\end{tabular} & \begin{tabular}[c]{@{}l@{}}``Loading a large script in the Rhino\\ debugger results in an endless loop\\ (100\% CPU utilization)"\end{tabular} & 16 \\ \hline
\textbf{\begin{tabular}[c]{@{}l@{}}Permission/\\ Deprecation\end{tabular}} & \begin{tabular}[c]{@{}l@{}}Bugs related to two main causes: (i) they are due to the presence, modification, or removal of \\deprecated method calls or APIs; (ii) problems related to unused API permissions are included\end{tabular} & \begin{tabular}[c]{@{}l@{}}``setTrackModification(boolean) not\\ deprecated; but does not work"\end{tabular} & 2 \\ \hline
\end{tabular}%
}
\vspace{-0.2cm}
\end{table*}

\subsection{Dataset}
We use the dataset of Parasaram et al.~\cite{parasaram2024factselectionproblemllmbased}, which includes 314 instances from the BugsInPy benchmark~\cite{10.1145/3368089.3417943}. Each instance includes the link to the fix commit that resolves the bug and the link to the buggy commit representing the repository state before the fix. This data contains bugs fixed by changes to a single function. 
Since our focus is on generating patches for APR, for bug localization we use \textit{Function-granular Perfect Fault Localization}, as recommended in prior work~\cite{parasaram2024factselectionproblemllmbased, 8730164}. %We aim to evaluate the LLM’s ability to generate fixes rather than locate bugs. 
%Therefore, we use the Function-granular Perfect Fault Localization approach as recommended in prior work~\cite{parasaram2024factselectionproblemllmbased, 8730164}.
However, LLMs will still have to identify the faulty lines within a buggy function.

We manually annotate the dataset using the bug categorization taxonomy proposed by Catolino et al.~\cite{CATOLINO2019165}. This taxonomy includes 9 categories: \textit{Configuration Issue}, \textit{Network Issue}, \textit{Database-Related Issue}, \textit{GUI-Related Issue}, \textit{Performance Issue}, \textit{Permission/Deprecation Issue}, \textit{Security Issue}, \textit{Program Anomaly Issue}, and \textit{Test Code-Related Issue}.
We use this taxonomy because it is comprehensive, capturing common root causes across diverse, real-world bugs (1,280 bug reports from 119 popular projects).
%This taxonomy was developed through a qualitative analysis of 1,280 bug reports from 119 popular projects to identify the root causes of reported bugs, making it well-suited for our study. 
The annotation instructions, including the definitions of each bug type and examples, are adapted from Catolino et al. and included in our replication package.

Two graduate students (each with 3+ years of experience in programming and qualitative research) conducted the annotation process. For each data point, we examined both textual and code-related data, including the \textit{GitHub issue thread}, \textit{pull request}, and the \textit{fix commit}. %By analyzing both textual and code-related clues, we determined the most appropriate category for each bug.
We followed an iterative analysis comprising multiple sessions. To ensure annotation reliability and mitigate bias, we employed Cohen’s Kappa inter-rater agreement~\cite{keppaArticle}.  In the first round, both annotators independently labeled an initial set of 50 data points, which yielded a Kappa score of 0.45. Follow-up discussions were conducted to refine the taxonomy's interpretation and resolve all disagreements. %In the second round, we re-annotated the same 50 data points and successfully resolved all conflicts. %but two conflicts. These two conflicts were resolved by including judgment from a third author.
In the second round, an additional 50 data points were annotated, which yielded a Kappa score of 0.8, indicating a high level of agreement~\cite{keppaArticle}. With this established consistency, the remaining 214 data points were divided between the two annotators and labeled independently.

Table \ref{table:dataset} presents the descriptions, examples, and distribution of bug types in our dataset. Six of the nine categories are represented; we did not find any \textit{Database-Related}, \textit{Security}, or \textit{Test Code-Related} bugs. Among the represented bug types, \textit{Program Anomaly} is the most frequent, while \textit{Permission/Deprecation} is the least.
Our dataset spans 16 well-known Python projects (e.g., \textit{pandas, youtube-dl, scrapy)}, %keras, luigi, thefuck, matplotlib, fastapi, tornado, black, ansible, tqdm, sanic, httpie, cookiecutter,} and \textit{PySnooper}), 
each with high star counts (ranging from 16k to 135k), contributor counts (25 to 463), and open issues (29 to 4172). 
%Out of the total 314 bugs, the projects \textit{pandas}, \textit{youtube-dl}, and \textit{scrapy} contain the highest number of bugs, with 107, 29, and 28, respectively. 
Full project-level statistics, including stars, contributors, open issues, and bug distributions, are provided in our replication package.

\subsection{Prompting Strategy}\label{prompt:layer1}

We use Chain-of-Thought prompting, a proven strategy for guiding LLMs to process context step by step, as demonstrated in APR and other tasks~\cite{parasaram2024factselectionproblemllmbased,10.1109/ASE56229.2023.00065, 10.1145/3664646.3664770, 10.1145/3690635}. 
We adopt a prompt structure similar to Parasaram et al.~\cite{parasaram2024factselectionproblemllmbased}, with a few key adjustments.
%Unlike their approach, which lacks a defined output format, 
We enforce standardized outputs by requiring all code to be wrapped in triple backticks. We also omit textual explanations to ensure the output contains only code.
%One major difference is that Parasaram et al. do not specify the format for the LLM's output. This lack of structure can lead to inconsistent results across different runs, complicating the extraction and testing of the corrected code in an automated setting. Additionally, they request that the LLMs provide reasoning behind the corrected code, resulting in a combined output of both code and explanation. Therefore, we enforce a standardized output format in our prompts, ensuring that all generated code is consistently wrapped in triple backticks. In addition, while we still ask the LLMs to analyze the code and identify the problems, we do not ask for any textual explanation within the output itself. This ensures that all outputs are uniformly structured. 
We tested our approach on 10 bug instances using GPT-4o-mini and Llama 3.3, running each bug 10 times per model. We achieved consistent outputs across all 200 runs.
Due to space constraints, we include the detailed prompt formats for each knowledge layer in our replication package. % is shown below:

\subsection{Research Questions and Information Types in Each Layer}

\noindent
\textbf{RQ1: To what extent can LLMs resolve different types of bugs using only immediate bug knowledge?}
%\subsubsection*{\underline{Information Types}}

Immediate bug knowledge is the contextual information directly related to the buggy code, including: \textit{Buggy Function}, \textit{Failing Tests}, \textit{Error Information}, \textit{Runtime Variables}, \textit{Angelic Variables}, %\textit{Buggy Class Declaration}, \textit{Used Method Signatures}, 
and \textit{GitHub Description}.
We incorporate them in the first layer of the knowledge injection pipeline, referred to as the \textbf{Bug Knowledge Layer}. %We exclude the \textit{Buggy Class Declaration} and \textit{Used Method Signatures} from this layer, as they introduce information beyond the immediate bug context.

The information types in this layer are adopted from Parasaram et al.~\cite{parasaram2024factselectionproblemllmbased}, and serve as a \textbf{baseline} for evaluating the impact of injecting other contextual information in subsequent layers (RQ2 and RQ3). 

%\subsubsection{Types of Information}
%In \textit{Bug Knowledge Layer}, we inject the following contextual information:

%\noindent
\textbf{Buggy Function.} This is the buggy code the LLM should fix. For each data instance, the function requiring repair is extracted using the fix commit (patch) provided by developers.

%\noindent
\textbf{Failing Tests.} Each bug instance also contains passing and failing tests. The buggy code can cause one or more tests to fail. Providing these failed tests could help the LLM understand the possible cause of the bug~\cite{10.1145/3650212.3680323}.

%\noindent
\textbf{Error Information.} We supply LLMs with error messages and stack traces generated by failing tests, as these can help models identify buggy code~\cite{10.1145/3650212.3680323, 52980}. 
%To obtain this information, we set up the appropriate environment for each bug by installing the required dependencies by cloning the project from GitHub and using Anaconda to install packages, reverting the repository to its state before the fix commit, and running the failing tests to capture error outputs.
To obtain this information, we first cloned each project from GitHub and reverted the repository to its state before the fix commit. We then used Anaconda to install the required dependencies and ran the failing tests to capture the corresponding error outputs.

%\noindent
\textbf{Runtime} and \textbf{Angelic Variables.} We extract two types of dynamic information: 
1) \textit{Runtime information}: values and types of function parameters and local variables during failing test execution; 2) \textit{Angelic forest}: the expected values that, if assigned to those variables during execution, would cause the failing test to pass. This behavioral data can improve LLM performance in bug fixing~\cite{parasaram2024factselectionproblemllmbased}. Prior work showed that such information supports synthesis-based program repair~\cite{7886945}. The values are extracted from correct program versions using instrumentation. Since runtime and expected values can be lengthy, especially when multiple test cases are available, we limit each bug instance to at most three input-output cases for both runtime and angelic variables.

%\noindent
\textbf{GitHub Description.} Providing textual descriptions of a bug can improve LLM-generated fixes~\cite{10.1145/3639478.3643526, ehsani2025detectingpromptknowledgegaps}. %Parasaram et al. traced each bug in their dataset to a corresponding GitHub Issue, but 
We extract the title and body description of issue reports associated with each bug in our dataset. We observed that 69 of 314 instances in the dataset do not have a linked report. To address this, we tried to extract the corresponding pull request. For 24 data points, no corresponding issue or pull request was found. We manually reviewed all linked issues and pull requests to ensure they only describe the bug and do not reveal the actual fix.

\vspace{-0.2cm}

\vspace{0.2cm}
\noindent
\textbf{RQ2: How does injecting repository-level knowledge enhance LLM-based repair of different bug types?}
%In this layer, we go beyond the immediate bug context by incorporating additional information from the repository. Specifically, we include information on the file containing the buggy function, related files, and relevant historical information about prior fixes. The goal is to help the LLM better understand the broader context of the bug, enabling it to generate more informed and accurate patches. We refer to this as the \textbf{Repository Knowledge Layer}. This layer is applied to bugs that remain unresolved after the \textit{Bug Knowledge Layer}, allowing us to evaluate how repository-level context contributes to improving repair outcomes. 
%\subsubsection{Types of Information}
%Each type of information is retrieved from the corresponding GitHub repository, using the state of the codebase before the introduction of the fix. 

%\subsubsection*{\underline{Information Types}}
In the \textbf{Repository Knowledge Layer}, we go beyond the immediate bug context by incorporating additional information from the repository: \textit{the buggy file, related files, and prior fix history}. Each type of information is retrieved from the state of the GitHub repository before the fix. We apply this knowledge injection to the bugs unresolved after the \textit{Bug Knowledge Layer}, allowing us to assess the impact of repository-level context on repair outcomes.

%\noindent
\textbf{Co-occurring Files.} %A repository’s commit history offers valuable insights into relationships between files~\cite{ajienka_empirical_2018, hrishikesh2025cochangegraphentropynew}. 
Prior APR research has shown that leveraging historical context can improve repair outcomes~\cite{10.1145/3395363.3397369, 7476644}. To identify files related to the one containing the buggy function, we analyze past commits and extract those that have most frequently been committed alongside the buggy file~\cite{ajienka_empirical_2018}. We refer to these as \textit{Co-occurring Files}. These files suggest a strong logical or functional connection, indicating that they often work together or depend on one another~\cite{hrishikesh2025cochangegraphentropynew}. Co-occurring files are not directly used in prompts. We use them as a heuristic to identify where the buggy function may be invoked elsewhere in the codebase, aiding in the discovery of structural dependencies discussed next.

%\noindent
\textbf{Structural Dependencies.} Understanding the context of the buggy function is critical for generating effective patches%~\cite{zhang2023neural%}, enabling models to reuse existing functions
~\cite{zhang2023neural, 10.1145/3597503.3639086, dikici_advancements_2025}. To capture structural dependencies, we examine both directions: the functions and classes invoked by the buggy function and those that invoke the buggy function elsewhere in the codebase. For each buggy function, we use Python’s AST and Jedi~\cite{noauthor_jedi_nodate} library to parse its body and extract referenced functions and classes. We then use the file’s import statements to resolve where those declarations originate and extract their definitions from the relevant files.
In the other direction, we identify functions and classes that invoke the buggy function. Since usages can be widespread and introduce noise especially for utility functions, we limit this search to the set of previously identified \textit{Co-occurring Files}. We then apply the same dependency resolution process to extract only the most relevant calling contexts.

%\noindent
\textbf{Latest Changes.} Including historical commit information related to the buggy function can improve the bug-fixing performance of LLMs~\cite{shi2025hafix}. Therefore, by analyzing the Git history of the repository, we extract the most recent commit made to the buggy function before the fix. This information suggests %This commit is then injected into the prompt to provide the LLM with valuable context about the latest changes and 
potential sources of error in the buggy function.

\vspace{0.2cm}
\noindent
\textbf{RQ3: How does injecting project-level knowledge further enhance LLM-based repair of different bug types?}
%\subsubsection*{\underline{Information Types}} 

In the %we inject project-level information to provide broader context about the software systems in which the bugs occur. We refer to this as the 
\textbf{Project Knowledge Layer}, %which builds on the previous layers by integrating both 
we inject \textit{project documentation and developer knowledge from past resolved issues}. We inject them to the unresolved bugs after the \textit{Repository Knowledge Layer}. The goal is to give the LLM a holistic view of the project, its constraints, and past bug-fixing strategies. %, enhancing its ability to generate more accurate patches. 
% An overview of our pipeline to extract this information is shown in Figure~\ref{fig:langchain}.

%\subsubsection{Types of Information}The following are the information added in \textit{Project Knowledge Layer}:

%\noindent
\textbf{Documentation.} Documentation can help LLMs infer code intent~\cite{wijaya2025readmellmframeworkhelpllms, 10.1145/3377811.3380430}. We extract information about buggy functions' usage, limitations, and behavior from the project’s official documentation.
%For each bug, we crawl and extract the documentation associated with the corresponding project. 
Since documentation can be extensive, we isolate the most relevant parts by applying a dense retrieval pipeline~\cite{10.1145/3611643.3616256, patil2025gitbugsbugreportsduplicate}. First, we split the documentation into equal-sized chunks and embed them using Sentence Transformer embeddings. We store the embeddings in a FAISS (Facebook AI Similarity Search)~\cite{douze2024faiss} vector index, enabling efficient similarity search against the buggy function. We use the full body of each buggy function, including the APIs it calls, as a query to retrieve the most relevant documentation chunks based on semantic similarity.
We then use LangChain’s QA Retrieval Agent~\cite{noauthor_langchain_nodate}, %powered by GPT-4o-mini, 
to query the embedded documents and extract four key pieces of information:
\textbf{(1)} What is the correct behavior of the \textit{buggy\_function}?
\textbf{(2)} What is the intended usage of the \textit{buggy\_function} and the APIs it uses? 
\textbf{(3)} What are the constraints, edge cases, or usage patterns for APIs used in \textit{buggy\_function}? 
\textbf{(4)} What are the limitations, warnings, or best practices of using APIs in the \textit{buggy\_function}? 

%\noindent
\textbf{Issue Resolution History.} %Building on the same intuition as documentation context, we also extract valuable insights from previously resolved issues within each project. Specifically, 
We focus on patched issues that occurred before the fix date of the target bug. These past issues often contain relevant knowledge about similar bugs and their resolutions~\cite{10.1145/3338906.3338935, motwani2023betterautomaticprogramrepair}.
To identify the most relevant issues, we apply dense retrieval to compare each past issue to the buggy function. We consider both the code  (patches) and the natural language (titles, descriptions, discussions, and labels). Inspired by previous studies~\cite{10.1145/3611643.3616256, patil2025gitbugsbugreportsduplicate}, we compute a similarity score for each dimension, textual and code, and combine them using an equal weighted average (50\% text, 50\% code). The issues, along with their metadata, are stored in a FAISS~\cite{douze2024faiss} vector index for efficient retrieval.
We then use LangChain’s QA Retrieval Agent, powered by GPT-4o-mini, to extract the following key insights:
\textbf{(1)} What relevant API behaviors, bugs, or patterns discussed in the past could inform fixing the \textit{buggy\_function}? \textbf{(2)} What concrete fix strategies were applied in similar scenarios to those in the \textit{buggy\_function}? \textbf{(3)} What specific code changes from past fixes could guide repairing the buggy function \textit{buggy\_function}? \textbf{(4)} What key developer insights, recommendations, or concerns discussed in past fixes are relevant to resolving the \textit{buggy\_function}? 

\begin{comment}
\subsubsection*{\underline{Prompt}}
This layer uses the same format as section \ref{prompt:layer2}), with added Project Knowledge Layer information as follows:
%Using the same logic and format used for the previous layer (section \ref{prompt:layer2}), the prompt is created with the injection of \textit{Project Knowledge Layer} information in the following structure:
\vspace{-0.2cm}
\begin{tcolorbox}[colback=blue!5!white,colframe=blue!35!black,boxsep=3pt,
left=3pt,
right=3pt,
top=3pt,
bottom=3pt]
\footnotesize
...[Previous Instructions]...

\textbf{\#\# Information retrieved from the documentation of the project related to the buggy function}

% \begin{verbatim}
\texttt{\{retrieved\_docs\}}
% \end{verbatim}

\textbf{\#\# Information retrieved from the past issue threads of the project similar to the buggy function}

% \begin{verbatim}
\texttt{\{retrieved\_issues\}}
% \end{verbatim}

\end{tcolorbox}
\end{comment}

%\subsubsection{Evaluation Measures}
%Similar to the previous layers, we use \textit{pass@k} and \textit{\#fixed} to measure the impact of \textit{Project Knowledge Layer} on LLM-based bug repair. For each bug that remained unresolved in the \textit{Repository Knowledge Layer}, we re-run them using the new prompts, generating $n = 10$ responses per instance as before.

\vspace{0.1cm}
\noindent
%\textbf{RQ4: Which types of bugs remain challenging for LLMs to solve, even after all layered contextual knowledge is injected?}
\textbf{RQ4: Which bug types remain challenging for LLMs, even after injecting all layers of contextual knowledge?}

We conduct an error analysis of the bugs that are unresolved after injecting all three layers of contextual knowledge. %For these bugs, we manually investigate the availability of the five types of information injected at the repository and project levels. This allows us to assess if some of these information are not available to be leveraged. 

First, for the unresolved bugs, we manually examine whether the five types of repository and project-level information were available and usable, since in real-world settings, such context may not always exist or be extractable.
%First, for the unresolved bugs at each layer, we manually examine whether the contextual information was available and usable, since in real-world settings, they may not always exist or be extractable.

Second, our goal is to identify the types and characteristics of the bugs that remain challenging for LLMs.
%, even when provided comprehensive context across bug, repository, and project levels. 
%we examine the injected repository and project-level context to understand why the added knowledge failed to help the model generate a correct fix. These two layers introduce five distinct types of contextual information. For each unresolved bug, We analyze how many of these elements were actually injected, allowing us to assess whether failures stem from missing context.
%We examine whether the unresolved bugs exhibit greater complexity compared to those that were successfully fixed 
To examine whether unresolved bugs are more complex than successfully fixed ones, we compute the following code complexity metrics using Python's widely-used Radon package~\cite{noauthor_radon_nodate, 10.1145/3487569, sandouka_python_2023}: 
%To do this, we compute several widely used code complexity metrics~\cite{5477581} for each buggy function, using Python's widely used radon package~\cite{noauthor_radon_nodate, 10.1145/3675888.3676139, 10.1145/3487569, sandouka_python_2023} as follows: 
a) \textbf{Cyclomatic Complexity}, which measures control flow complexity based on decision points; b) \textbf{Maintainability Index}, a composite metric reflecting how easily the code can be understood and maintained; c) \textbf{Lines of Code (LOC)}, capturing function length; and d) \textbf{Halstead Metrics} (Volume, Difficulty, and Effort), which reflect the cognitive complexity based on operators and operands. 
%We compare the distributions of these metrics between resolved and unresolved bugs to identify whether code complexity is associated with degraded LLM performance. 
As we are comparing two groups (resolved vs. unresolved bugs) on various continuous metrics, we run statistical tests to assess the significance of differences. Specifically, as the data does not follow a normal distribution (verified using the Shapiro–Wilk test), we use the non-parametric Mann–Whitney U test~\cite{ranganathan_introduction_2021}. Additionally, we compute Cohen's d~\cite{sullivan_using_2012} to measure the effect size and indicate which group tends to exhibit higher complexity across each metric.
This analysis helps us assess whether unresolved bugs may require more sophisticated reasoning or deeper understanding, potentially posing difficulties to current LLMs.

% We begin by examining the unresolved bugs across different bug types for both models. For each case, we analyze the patches generated in each contextual layer to determine why the LLM failed to produce a valid fix. Specifically, we categorize failed patches into two broad outcomes: \textbf{Test Failures}: The patch was syntactically valid but failed the test suite, indicating that the generated solution did not meet the expected behavioral criteria, and \textbf{Errored Patches}: The patch caused runtime or static errors, such as compilation issues or invalid API calls. For errored patches, we further classify them based on established taxonomies of LLM-generated code errors~\cite{llm_errors}: \textbf{Hallucinations}: The patch includes fabricated elements, such as non-existent APIs, functions, or incorrect usage of existing libraries, and \textbf{Syntax/Logical Errors}: The patch is syntactically incorrect or structurally flawed in a way that causes runtime or logical failures.
% We analyze these failure modes to see whether the type of mistake changes across layers, and whether contextual information reduces hallucinations or syntax errors, even if the bug remains unresolved. This helps us understand not just whether a bug is fixed, but how the model’s reasoning changes or, fails to change, when additional context is injected.

\subsection{Baselines}
We compare our approach against two baselines: (1) Parasaram et al.~\cite{parasaram2024factselectionproblemllmbased}'s approach, which leverages various bug-related contextual information to produce patches; and (2) an approach where information from all three layers (bug, repository, and project knowledge) is injected simultaneously for every bug. The second baseline allows us to assess whether progressive, hierarchical injection is more effective than providing all context at once.

\subsection{Evaluation Metrics}
The nondeterminism of LLMs can lead to varying outcomes across responses, which presents challenges when analyzing results. To address this, we evaluate the generated responses using the \textit{pass@k} measure, which represents the probability that at least one out of k queries successfully resolves the problem.
% \vspace{-0.3cm}
\begin{equation}\label{form:1}
    \text{Pass@k} = \mathbb{E}_Q \left[1 - \frac{\binom{n - C}{k}}{\binom{n}{k}} \right]
\end{equation}
Here, $\mathbb{E}_Q$ denotes the expectation over the set of LLM responses for the set of queries (prompts) $Q$, $n$ is the total number of responses obtained from the LLM where $n>k$, and $C$ is the number of successful responses among them. For our program repair task, we consider a response successful if the extracted patch satisfies a correctness criterion, which we approximate by passing the test suite. We choose $n=10$ and report Pass@k with k=1, 3, 5, following prior studies~\cite{shi2025hafixhistoryaugmentedlargelanguage, parasaram2024factselectionproblemllmbased}.

We also report the number of fixed bugs in our dataset (\textit{\#fixed}). This is a common metric for evaluating APR tools that iteratively generate and test patches, indicating the number of bugs for which at least one generated patch passes all tests~\cite{parasaram2024factselectionproblemllmbased, 10.1145/2771783.2771791}. The \#fixed is measured by: 
% \vspace{-0.1cm}
\begin{equation}\label{form:2}
    \#\text{fixed}\ (\text{LLM} (J)) \triangleq \left| \{ b \mid j \in J, C_j > 0 \} \right|
\end{equation}
where \( j = (b, F) \) represents a prompt, and \( C_j \) is the number of responses that pass the tests for the given prompt.
All plausible fixes are tested and manually verified to be the correct fix.
%For each bug that remained unresolved in the previous layer, we re-run them using the new prompts, generating $n = 10$ responses per instance.

%% file: 04_results.tex
\section{Results and Discussion}
    \subsection{RQ1: To what extent can LLMs resolve different types of bugs using only immediate bug knowledge?}
\textbf{Results.}
Table~\ref{tab:overall_results} presents the results of running our entire dataset through the \textit{Bug Knowledge Layer} using GPT-4o-mini and Llama 3.3. When provided only with local context surrounding the buggy function, GPT-4o-mini resolves 197 out of 314 bugs (62\%), while Llama slightly outperforms it by fixing 207 bugs (65\%).
In terms of \textit{Pass@k} performance, both models show consistent capabilities, though Llama is more reliable overall. Llama achieves a \textit{Pass@1} of 47\%, compared to 38\% for GPT-4o-mini, indicating a 9\% higher chance of generating a correct fix on the first attempt. As \textit{k} increases, performance improves for both models: \textit{Pass@3} and \textit{Pass@5} exceed 50\%, with Llama reaching 61\% at \textit{Pass@5} and maintaining a consistent lead across all values of \textit{k}.

Table~\ref{tab:overall_results_bug_type} breaks down performance by bug type. Llama performs better on \textit{Network}  (e.g., Llama 3.3 resolves 72\% bugs compared to 68\% by GPT-4o-mini), \textit{Program Anomaly}, \textit{Configuration}, and \textit{Permission/Deprecation} bugs, while GPT-4o-mini shows an advantage for \textit{Performance} bugs. For \textit{GUI}-related bugs, both models fix 16 out of 28 instances (57\%), with 13 of those fixes overlapping and 3 unique to each model.

%\noindent
\textbf{Discussion.} Since some of our metrics in this layer are adopted from Parasaram et al.~\cite{parasaram2024factselectionproblemllmbased}, we compare our results to theirs. They achieved a total fix rate of 56\% (177/314 bugs), using GPT-3.5 Turbo and Llama3-70B. Our best-performing model in this layer, Llama 3.3, achieves statistically significant improvements over Parasaram et al.'s results, supported by a Chi-square test (\textit{p-value}=0.017) and Z-test (\textit{p-value}=0.014). 
We exclude two of Parasaram et al.'s metrics, \textit{Buggy Class Declaration} and \textit{Used Method Signatures}, as they introduce information beyond the immediate bug context.
%This suggests that injecting all available bug-related information, rather than structurally injecting it based on knowledge layers of different granularity, could lead to reduced performance. 
The results indicate the strength of our layered approach. Bugs that can be resolved with only bug knowledge are fixed early, without the risk of confusing the model with unnecessary context, while subsequent layers will progressively address bugs that require a deeper understanding of the repository or project (as we will later find out in RQ2 and RQ3). The use of newer LLMs with larger context windows (128k vs. 8k) further supports this approach by enabling the injection of richer knowledge without exceeding model limitations. 
%Our findings offer two key observations:
%\textbf{(1)} Newer models (e.g., Llama3.3 over Llama3) are more capable of processing large-scale context, leading to improved APR performance. Thus, 
%\textbf{(2)} the challenge is shifting from \textit{selecting} useful information to strategically extracting and \textit{injecting} only relevant information into the prompt, without confusing the model with noisy data. 
%\textbf{(2)} LLMs have become more capable of processing and utilizing large-scale contextual input.
%Given the performance improvement on using newer models (e.g., Llama3.3 over Llama3) with similar contextual information, LLMs have become more capable in APR tasks.

Additionally, our bug-type analysis provides actionable guidance for model selection under limited context. For instance, if working with a \textit{Performance} bug and restricted to bug-level information (e.g., error logs, test failures), GPT-4o-mini may be a better choice. In contrast, for bug types like \textit{Program Anomaly}, Llama shows stronger results. This indicates that the effectiveness of each model can vary depending on the type of bug and the scope of available context.
Across all bug types, the Llama model consistently achieves higher \textit{Pass@k} scores, suggesting more reliable performance within fewer attempts compared to GPT-4o-mini.
Finally, Llama failed to fix 107 bugs and GPT-4o-mini failed on 117, with 78 overlapping. This intersection suggests that there are particularly challenging cases that LLMs are unable to fix with immediate bug context alone.
% This intersection suggests consistent LLM blind spots or particularly challenging cases that may require broader context.
%Finally, among the unresolved bugs in this layer, both models failed to fix the same 78 bugs, suggesting consistent blind spots. Llama failed to fix 107 bugs, GPT-4o-mini failed on 117, and 78 bugs overlapped, highlighting challenging cases that require broader and more contextual information.

\begin{table}[t!]
\caption{Performance of Knowledge Injection Layers for each LLM}
\vspace{-0.2cm}
\label{tab:overall_results}
\resizebox{\columnwidth}{!}{%
\renewcommand{\arraystretch}{1.2}
\begin{tabular}{|p{2.5cm}|l|c|c|c|c|}
\hline
\textbf{Knowledge Injection} & \textbf{LLM} & \textbf{Pass@1} & \textbf{Pass@3} & \textbf{Pass@5} & \textbf{\begin{tabular}[c]{@{}c@{}}\%Fixed\\ Bugs\end{tabular}} \\ \hline
\textbf{Bug Knowledge} & Llama 3.3 & 0.47 & 0.58 & 0.61 & 65\% (207/314) \\ \cline{2-6} 
\textbf{Layer} & GPT-4o-mini & 0.38 & 0.50 & 0.56 & 62\% (197/314) \\ \hline
\textbf{Repository} & Llama 3.3 & 0.50 & 0.63 & 0.68 & 74\% (235/314) \\ \cline{2-6} 
\textbf{Knowledge Layer} & GPT-4o-mini & 0.40 & 0.54 & 0.61 & 70\% (221/314) \\ \hline
\textbf{Project Knowledge} & Llama 3.3 & \textbf{0.50} & \textbf{0.65} & \textbf{0.71} & \textbf{79\% (250/314)} \\ \cline{2-6} 
\textbf{Layer} & GPT-4o-mini & 0.40 & 0.55 & 0.62 & 73\% (229/314) \\ \hline\hline
\textbf{All Knowledge} & Llama 3.3 & 0.38 & 0.53 & 0.59 & 65\% (207/314) \\ \cline{2-6} 
\textbf{Layers at Once} & GPT-4o-mini & 0.27 & 0.37 & 0.41 & 48\% (152/314) \\ \hline
\end{tabular}%
}
\vspace{-0.2cm}
\end{table}

\begin{table*}[t!]
\caption{Performance of Knowledge Injection Layers by Bug Type}
\vspace{-0.2cm}
\label{tab:overall_results_bug_type}
\resizebox{\textwidth}{!}{%
\begin{tabular}{|l|ccccccccc|ccccccccc|}
\hline
{\textbf{Bug Type\textbackslash{}LLM}} & \multicolumn{9}{c|}{\textbf{GPT-4o-mini}} & \multicolumn{9}{c|}{\textbf{Llama 3.3}} \\ \cline{2-19} 
 & \multicolumn{3}{c|}{\textbf{\begin{tabular}[c]{@{}c@{}}Bug\\ Knowledge Layer\end{tabular}}} & \multicolumn{3}{c|}{\textbf{\begin{tabular}[c]{@{}c@{}}Repository\\ Knowledge Layer\end{tabular}}} & \multicolumn{3}{c|}{\textbf{\begin{tabular}[c]{@{}c@{}}Project\\ Knowledge Layer\end{tabular}}} & \multicolumn{3}{c|}{\textbf{\begin{tabular}[c]{@{}c@{}}Bug\\ Knowledge Layer\end{tabular}}} & \multicolumn{3}{c|}{\textbf{\begin{tabular}[c]{@{}c@{}}Repository\\ Knowledge Layer\end{tabular}}} & \multicolumn{3}{c|}{\textbf{\begin{tabular}[c]{@{}c@{}}Project \\ Knowledge Layer\end{tabular}}} \\ \cline{2-19} 
 & \multicolumn{1}{c|}{\textbf{\begin{tabular}[c]{@{}c@{}}AVG\\ Pass@1\end{tabular}}} & \multicolumn{1}{c|}{\textbf{\begin{tabular}[c]{@{}c@{}}AVG\\ Pass@3\end{tabular}}} & \multicolumn{1}{c|}{\textbf{\begin{tabular}[c]{@{}c@{}}\%Fixed\\ Bugs\end{tabular}}} & \multicolumn{1}{c|}{\textbf{\begin{tabular}[c]{@{}c@{}}AVG\\ Pass@1\end{tabular}}} & \multicolumn{1}{c|}{\textbf{\begin{tabular}[c]{@{}c@{}}AVG\\ Pass@3\end{tabular}}} & \multicolumn{1}{c|}{\textbf{\begin{tabular}[c]{@{}c@{}}\%Fixed\\ Bugs\end{tabular}}} & \multicolumn{1}{c|}{\textbf{\begin{tabular}[c]{@{}c@{}}AVG\\ Pass@1\end{tabular}}} & \multicolumn{1}{c|}{\textbf{\begin{tabular}[c]{@{}c@{}}AVG\\ Pass@3\end{tabular}}} & \textbf{\begin{tabular}[c]{@{}c@{}}\%Fixed\\ Bugs\end{tabular}} & \multicolumn{1}{c|}{\textbf{\begin{tabular}[c]{@{}c@{}}AVG\\ Pass@1\end{tabular}}} & \multicolumn{1}{c|}{\textbf{\begin{tabular}[c]{@{}c@{}}AVG\\ Pass@3\end{tabular}}} & \multicolumn{1}{c|}{\textbf{\begin{tabular}[c]{@{}c@{}}\%Fixed\\ Bugs\end{tabular}}} & \multicolumn{1}{c|}{\textbf{\begin{tabular}[c]{@{}c@{}}AVG\\ Pass@1\end{tabular}}} & \multicolumn{1}{c|}{\textbf{\begin{tabular}[c]{@{}c@{}}AVG\\ Pass@3\end{tabular}}} & \multicolumn{1}{c|}{\textbf{\begin{tabular}[c]{@{}c@{}}\%Fixed\\ Bugs\end{tabular}}} & \multicolumn{1}{c|}{\textbf{\begin{tabular}[c]{@{}c@{}}AVG\\ Pass@1\end{tabular}}} & \multicolumn{1}{c|}{\textbf{\begin{tabular}[c]{@{}c@{}}AVG\\ Pass@3\end{tabular}}} & \textbf{\begin{tabular}[c]{@{}c@{}}\%Fixed\\ Bugs\end{tabular}} \\ \hline
\textbf{\begin{tabular}[c]{@{}l@{}}Program\\ Anomaly\end{tabular}} & \multicolumn{1}{c|}{0.34} & \multicolumn{1}{c|}{0.47} & \multicolumn{1}{c|}{\begin{tabular}[c]{@{}c@{}}60\%\\ (113/187)\end{tabular}} & \multicolumn{1}{c|}{0.35} & \multicolumn{1}{c|}{0.50} & \multicolumn{1}{c|}{\begin{tabular}[c]{@{}c@{}}67\%\\ (126/187)\end{tabular}} & \multicolumn{1}{c|}{0.35} & \multicolumn{1}{c|}{0.50} & \begin{tabular}[c]{@{}c@{}}68\%\\ (129/187)\end{tabular} & \multicolumn{1}{c|}{0.41} & \multicolumn{1}{c|}{0.53} & \multicolumn{1}{c|}{\begin{tabular}[c]{@{}c@{}}62\%\\ (117/187)\end{tabular}} & \multicolumn{1}{c|}{0.43} & \multicolumn{1}{c|}{0.57} & \multicolumn{1}{c|}{\begin{tabular}[c]{@{}c@{}}70\%\\ (132/187)\end{tabular}} & \multicolumn{1}{c|}{\textbf{0.43}} & \multicolumn{1}{c|}{\textbf{0.59}} & \textbf{\begin{tabular}[c]{@{}c@{}}76\%\\ (142/187)\end{tabular}} \\ \hline
\textbf{Network} & \multicolumn{1}{c|}{0.43} & \multicolumn{1}{c|}{0.55} & \multicolumn{1}{c|}{\begin{tabular}[c]{@{}c@{}}68\%\\ (32/47)\end{tabular}} & \multicolumn{1}{c|}{0.48} & \multicolumn{1}{c|}{0.62} & \multicolumn{1}{c|}{\begin{tabular}[c]{@{}c@{}}78\%\\ (37/47)\end{tabular}} & \multicolumn{1}{c|}{0.48} & \multicolumn{1}{c|}{0.63} & \textbf{\begin{tabular}[c]{@{}c@{}}80\%\\ (38/47)\end{tabular}} & \multicolumn{1}{c|}{0.53} & \multicolumn{1}{c|}{0.65} & \multicolumn{1}{c|}{\begin{tabular}[c]{@{}c@{}}72\%\\ (34/47)\end{tabular}} & \multicolumn{1}{c|}{0.57} & \multicolumn{1}{c|}{0.69} & \multicolumn{1}{c|}{\begin{tabular}[c]{@{}c@{}}76\%\\ (36/47)\end{tabular}} & \multicolumn{1}{c|}{\textbf{0.57}} & \multicolumn{1}{c|}{\textbf{0.71}} & \begin{tabular}[c]{@{}c@{}}78\%\\ (37/47)\end{tabular} \\ \hline
\textbf{GUI} & \multicolumn{1}{c|}{0.30} & \multicolumn{1}{c|}{0.40} & \multicolumn{1}{c|}{\begin{tabular}[c]{@{}c@{}}57\%\\ (16/28)\end{tabular}} & \multicolumn{1}{c|}{0.32} & \multicolumn{1}{c|}{0.46} & \multicolumn{1}{c|}{\begin{tabular}[c]{@{}c@{}}64\%\\ (18/28)\end{tabular}} & \multicolumn{1}{c|}{0.37} & \multicolumn{1}{c|}{0.52} & \begin{tabular}[c]{@{}c@{}}75\%\\ (21/28)\end{tabular} & \multicolumn{1}{c|}{0.44} & \multicolumn{1}{c|}{0.53} & \multicolumn{1}{c|}{\begin{tabular}[c]{@{}c@{}}57\%\\ (16/28)\end{tabular}} & \multicolumn{1}{c|}{0.51} & \multicolumn{1}{c|}{0.64} & \multicolumn{1}{c|}{\begin{tabular}[c]{@{}c@{}}71\%\\ (20/28)\end{tabular}} & \multicolumn{1}{c|}{\textbf{0.51}} & \multicolumn{1}{c|}{\textbf{0.70}} & \textbf{\begin{tabular}[c]{@{}c@{}}82\%\\ (23/28)\end{tabular}} \\ \hline
\textbf{Configuration} & \multicolumn{1}{c|}{0.48} & \multicolumn{1}{c|}{0.60} & \multicolumn{1}{c|}{\begin{tabular}[c]{@{}c@{}}70\%\\ (24/34)\end{tabular}} & \multicolumn{1}{c|}{0.52} & \multicolumn{1}{c|}{0.65} & \multicolumn{1}{c|}{\begin{tabular}[c]{@{}c@{}}76\%\\ (26/34)\end{tabular}} & \multicolumn{1}{c|}{0.52} & \multicolumn{1}{c|}{0.65} & \begin{tabular}[c]{@{}c@{}}76\%\\ (26/34)\end{tabular} & \multicolumn{1}{c|}{0.68} & \multicolumn{1}{c|}{0.77} & \multicolumn{1}{c|}{\begin{tabular}[c]{@{}c@{}}85\%\\ (29/34)\end{tabular}} & \multicolumn{1}{c|}{0.70} & \multicolumn{1}{c|}{0.81} & \multicolumn{1}{c|}{\begin{tabular}[c]{@{}c@{}}94\%\\ (32/34)\end{tabular}} & \multicolumn{1}{c|}{\textbf{0.70}} & \multicolumn{1}{c|}{\textbf{0.82}} & \textbf{\begin{tabular}[c]{@{}c@{}}97\%\\ (33/34)\end{tabular}} \\ \hline
\textbf{Performance} & \multicolumn{1}{c|}{0.50} & \multicolumn{1}{c|}{0.64} & \multicolumn{1}{c|}{\begin{tabular}[c]{@{}c@{}}75\%\\ (12/16)\end{tabular}} & \multicolumn{1}{c|}{0.53} & \multicolumn{1}{c|}{0.70} & \multicolumn{1}{c|}{\begin{tabular}[c]{@{}c@{}}81\%\\ (13/16)\end{tabular}} & \multicolumn{1}{c|}{0.53} & \multicolumn{1}{c|}{0.70} & \begin{tabular}[c]{@{}c@{}}81\%\\ (13/16)\end{tabular} & \multicolumn{1}{c|}{0.51} & \multicolumn{1}{c|}{0.60} & \multicolumn{1}{c|}{\begin{tabular}[c]{@{}c@{}}62\%\\ (10/16)\end{tabular}} & \multicolumn{1}{c|}{\textbf{0.58}} & \multicolumn{1}{c|}{\textbf{0.74}} & \multicolumn{1}{c|}{\textbf{\begin{tabular}[c]{@{}c@{}}87\%\\ (14/16)\end{tabular}}} & \multicolumn{1}{c|}{0.58} & \multicolumn{1}{c|}{0.74} & \begin{tabular}[c]{@{}c@{}}87\%\\ (14/16)\end{tabular} \\ \hline
\textbf{\begin{tabular}[c]{@{}l@{}}Permission/\\ Deprecation\end{tabular}} & \multicolumn{1}{c|}{0.0} & \multicolumn{1}{c|}{0.0} & \multicolumn{1}{c|}{\begin{tabular}[c]{@{}c@{}}0\%\\ (0/2)\end{tabular}} & \multicolumn{1}{c|}{0.10} & \multicolumn{1}{c|}{0.26} & \multicolumn{1}{c|}{\begin{tabular}[c]{@{}c@{}}50\%\\ (1/2)\end{tabular}} & \multicolumn{1}{c|}{0.10} & \multicolumn{1}{c|}{0.26} & \begin{tabular}[c]{@{}c@{}}50\%\\ (1/2)\end{tabular} & \multicolumn{1}{c|}{\textbf{0.40}} & \multicolumn{1}{c|}{\textbf{0.50}} & \multicolumn{1}{c|}{\textbf{\begin{tabular}[c]{@{}c@{}}50\%\\ (1/2)\end{tabular}}} & \multicolumn{1}{c|}{0.40} & \multicolumn{1}{c|}{0.50} & \multicolumn{1}{c|}{\begin{tabular}[c]{@{}c@{}}50\%\\ (1/2)\end{tabular}} & \multicolumn{1}{c|}{0.40} & \multicolumn{1}{c|}{0.50} & \begin{tabular}[c]{@{}c@{}}50\%\\ (1/2)\end{tabular} \\ \hline
\textbf{Overall} & \multicolumn{1}{c|}{0.38} & \multicolumn{1}{c|}{0.50} & \multicolumn{1}{c|}{\begin{tabular}[c]{@{}c@{}}62\%\\ (197/314)\end{tabular}} & \multicolumn{1}{c|}{0.40} & \multicolumn{1}{c|}{0.54} & \multicolumn{1}{c|}{\begin{tabular}[c]{@{}c@{}}70\%\\ (221/314)\end{tabular}} & \multicolumn{1}{c|}{0.40} & \multicolumn{1}{c|}{0.55} & \begin{tabular}[c]{@{}c@{}}73\%\\ (229/314)\end{tabular} & \multicolumn{1}{c|}{0.47} & \multicolumn{1}{c|}{0.58} & \multicolumn{1}{c|}{\begin{tabular}[c]{@{}c@{}}65\%\\ (207/314)\end{tabular}} & \multicolumn{1}{c|}{0.50} & \multicolumn{1}{c|}{0.63} & \multicolumn{1}{c|}{\begin{tabular}[c]{@{}c@{}}74\%\\ (235/314)\end{tabular}} & \multicolumn{1}{c|}{\textbf{0.50}} & \multicolumn{1}{c|}{\textbf{0.65}} & \textbf{\begin{tabular}[c]{@{}c@{}}79\%\\ (250/314)\end{tabular}} \\ \hline
\end{tabular}%
}
\vspace{-0.2cm}
\end{table*}

\begin{comment}
\begin{figure}[h]
    \centering
    \includegraphics[width=0.47\textwidth]{figs/Frame 12.pdf}
    \vspace{-0.2cm}
    \caption{Comparison of Patch Generation with Repository Knowledge Layer Injected to LLMs. Bug \#36 in Project Keras}
    \label{fig:keras36}
\end{figure}

% \begin{figure}[h]
%     \centering
%     \begin{subfigure}[b]{0.45\textwidth}
%         \centering
%         \includegraphics[width=\textwidth]{figs/Frame 12.pdf}
%         \caption{Bug \#36 in Project Keras}
%         \label{fig:keras36}
%     \end{subfigure}
%     \hfill
%     \begin{subfigure}[b]{0.45\textwidth}
%         \centering
%         \includegraphics[width=\textwidth]{figs/Frame 13.pdf}
%         \caption{Bug \#116 in Project Pandas}
%         \label{fig:pandas116}
%     \end{subfigure}
%     \caption{Comparison of Patch Generation with Repository Knowledge Layer Injected to LLMs}
%     \label{fig:repo_combined}
% \end{figure}
\end{comment}

\subsection{RQ2: How does injecting repository-level knowledge enhance LLM-based repair of different bug types?}
\textbf{Results.}
We re-patch the previously unresolved bugs from the \textit{Bug Knowledge Layer} (107 for Llama and 117 for GPT-4o-mini) using updated prompts that inject additional information from the \textit{Repository Knowledge Layer}. The results in Table~\ref{tab:overall_results} show substantial improvements for GPT-4o-mini and a statistically significant improvement for Llama 3.3.
GPT-4o-mini successfully resolves 24 additional bugs, increasing its overall fix rate to \textbf{70\%} (Chi-square \textit{p-value}=0.051 and Z-test \textit{p-value}=0.042), while Llama 3.3 resolves 28 more, reaching \textbf{74\%} (Chi-square \textit{p-value}=0.018 and Z-test \textit{p-value}=0.014). This represents a significant performance gain of \textbf{8–9\%} when repository-specific context is layered on top of local bug knowledge. Both models also show consistent improvements in their \textit{Pass@k} scores, with Llama 3.3 continuing to outperform GPT-4o-mini in overall consistency, surpassing the 50\% threshold for \textit{Pass@1} in this layer.

As shown in Table~\ref{tab:overall_results_bug_type}, both models also demonstrate performance improvements across all bug types in the \textit{Repository Knowledge Layer} compared to their respective results in the \textit{Bug Knowledge Layer}. GPT-4o-mini resolves 13 additional \textit{Program Anomaly}, 5 \textit{Network}, 2 \textit{GUI}, 2 \textit{Configuration}, 1 \textit{Performance}, and 1 \textit{Permission/Deprecation} bugs. In comparison, Llama fixes 15 more \textit{Program Anomaly}, 2 \textit{Network}, 4 \textit{GUI}, 3 \textit{Configuration}, and 4 \textit{Performance} bugs, but sees no improvement for \textit{Permission/Deprecation} bugs. It is to be noted that there are only 2 \textit{Permission/Deprecation} bugs in our dataset.
This indicates that, regardless of the bug type, adding repository-level context consistently benefits LLM-based repair. While GPT-4o-mini sees greater gains on the number of fixed \textit{Network} bugs, Llama shows stronger improvements overall, outperforming the GPT model across different types of bugs in terms of fix rate and \textit{Pass@k}.

% Relying solely on the local scope of the buggy function is often insufficient for understanding the broader context needed to generate an appropriate fix. By incorporating structural dependencies, related files, and commit history, this layer equips LLMs with a deeper understanding of the codebase, enabling more fixes across a variety of bug types. In addition, Llama still proves to be the more consistent model with higher improvements in \textit{Pass@k} measures across different types of bugs.

%\noindent
\textbf{Discussion.} These results highlight the role of repository-level knowledge, such as related files, structural dependencies, and recent commit history, in enhancing LLM-based bug repair. When local bug knowledge is insufficient, this additional context allows the model to better understand how the buggy function fits within the broader codebase.
%, leading to more accurate and informed patches.
The fact that both models improve across bug types suggests that repository knowledge is generically beneficial.
Moreover, Llama 3.3 continues to show stronger consistency in this layer, with higher \textit{Pass@k} scores across all bug types. This suggests that it may be better equipped to take advantage of deeper contextual code information when available compared to GPT-4o-mini.

\begin{figure}[h]
    \centering
    \includegraphics[width=0.48\textwidth]{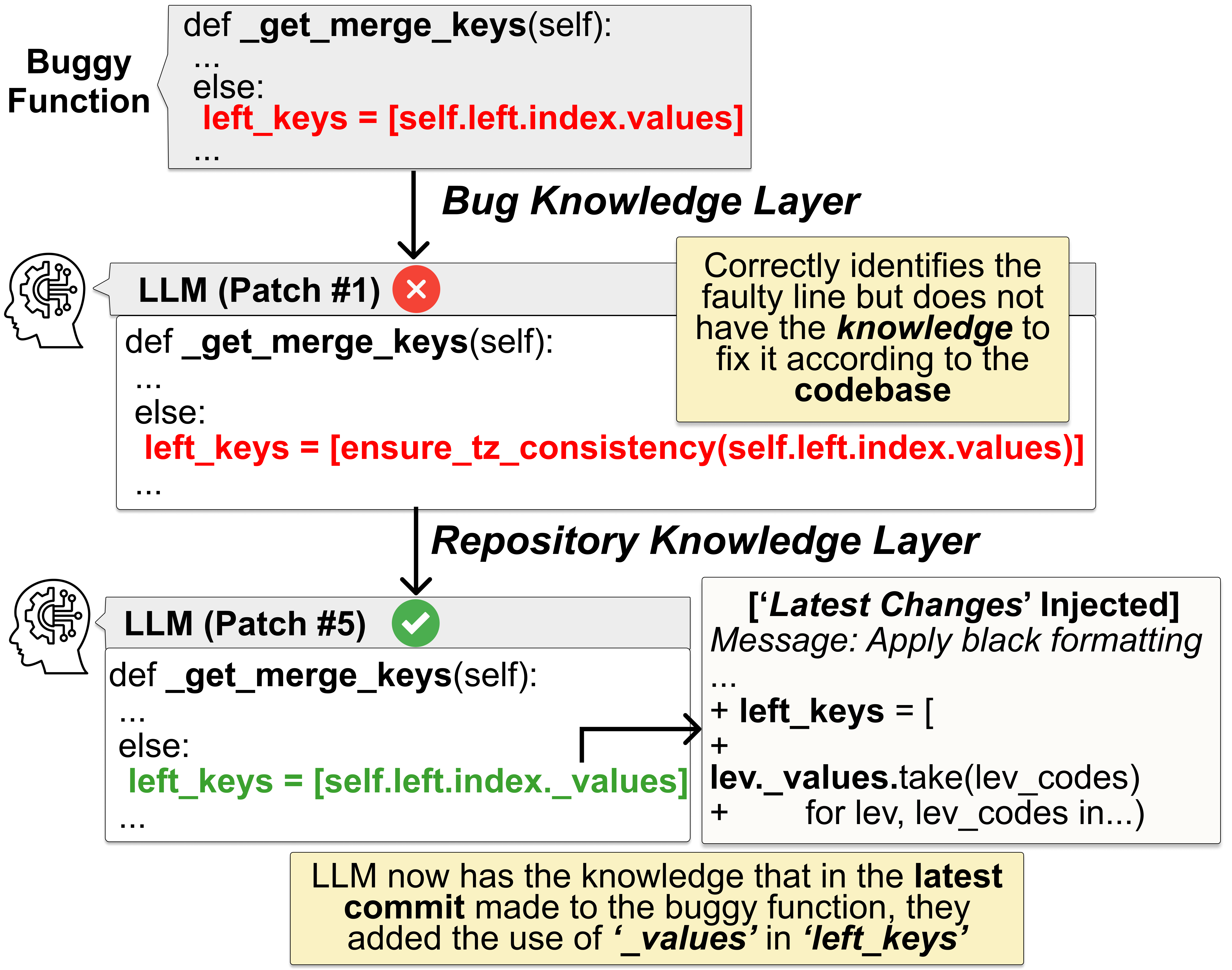}
    \vspace{-0.3cm}
    \caption{Comparison of Patch Generation with Repository Knowledge Layer Injected to LLMs. Bug \#116 in Project Pandas}
    \label{fig:pandas116}
    % \vspace{-0.2cm}
\end{figure}

To better understand how repository-level information contributes to bug repair, we manually analyzed the additional bugs solved in this layer (28 by Llama and 24 by GPT-4o-mini). For every fixed bug, we examined the structure and logic of the buggy function, the additional context retrieved at the current layer, the generated patch and corresponding test outcomes, and then assessed whether the added context plausibly enabled the fix. Manual analysis was done by the first author, and findings were refined through collaborative thematic analysis with the other authors during group discussions, from which we derive the reported patterns and insights.
We found that structural dependencies, related files, and commit history provide crucial missing context that helps LLMs infer the intended behavior of the code and produce accurate fixes. For example (Figure~\ref{fig:pandas116}), a \textit{Program Anomaly} bug from pandas~\cite{noauthor_bug_pandas_116} required changing ``\textit{index.values}" to ``\textit{index.\_values}" in one of the lines. In the \textit{Bug Knowledge Layer}, Llama was unaware of ``\textit{\_values}" as a valid attribute due to its absence in the local context. However, the structural dependencies and the latest commit affecting the function added additional relevant information, and Llama was able to understand the correct usage and generate the right patch.
%In another case, bug \#166 in pandas classified as a \textit{Program Anomaly}~\cite{noauthor_bug_pandas_166}, involved a misconfiguration in the use of the ``\textit{concat()}" function. The correct fix was simply adding ``\textit{sort=sort}" as a parameter to this function. Without structural information, Llama failed to recognize that this was a valid argument. Injecting the full definition of the ``\textit{concat()}" revealed its parameter structure, enabling the model to generate correct fixes.

These patterns extend to other bug types as well. In a \textit{Performance} bug in the luigi project~\cite{noauthor_merge_luigi}, the solution involved setting ``\textit{record\_task\_history=False}" in a call to ``\textit{CentralPlannerScheduler}". In the previously generated patches, the LLMs tried to guess what the right call to the class is, even hallucinating non-existent variables such as ``\textit{self.\_config}". Once additional knowledge like the ``\textit{CentralPlannerScheduler}" class was injected into the prompt through structural dependency analysis, the LLMs were able to generate correct patches. 
In a \textit{Network} bug from the fastapi project~\cite{noauthor_bug_fastapi_12}, injecting the definition of the ``\textit{HTTPException}" class likely enabled GPT-4o-mini to understand its internal structure and generate correct patches that handle the network error properly. 
Similarly, in a \textit{GUI} bug from matplotlib~\cite{noauthor_bug_mat_30}, the addition of usage examples from co-occurring files along with recent changes to the buggy function provided additional clues about its correct behavior and helped the model identify the underlying issue.

%Another notable case is a \textit{Program Anomaly} bug (\#36) from the Keras project~\cite{noauthor_bug_keras_36}. The issue stemmed from mismatched tensor dimensions in the separable convolution functionality of TensorFlow. By injecting relevant functions such as ``\textit{\_preprocess\_conv1d\_input()}", ``\textit{expand\_dims()}", and ``\textit{squeeze()}", all of which were used within the buggy function, the GPT model gained sufficient context to generate valid patches (see Figure~\ref{fig:keras36}).

Collectively, these cases show that effective LLM-guided bug repair often needs more than isolated code snippets; it requires situating the buggy function within the larger repository context. Structural and historical insights can provide the missing links that allow LLMs to reason more effectively about how to fix buggy code across different bug types.

\subsection{RQ3: How does injecting project-level knowledge further enhance LLM-based repair of different bug types?}
\textbf{Results.}
The third row of Table~\ref{tab:overall_results} shows the results after fixing the unresolved bugs after the \textit{Repository Knowledge Layer} using additional project-level knowledge in the \textit{Project Knowledge Layer} (79 unresolved bugs for Llama and 93 for GPT). With this final layer of knowledge injection, Llama 3.3 successfully fixes 15 more bugs, increasing its overall fix rate to \textbf{79\%}, \textbf{5\%} more than the previous layer. GPT-4o-mini shows a \textbf{3\%} improvement, fixing 7 additional bugs and reaching a \textbf{73\%} fix rate.
Both models show marginal gains in \textit{Pass@1}, while improvements are more noticeable in \textit{Pass@3} and \textit{Pass@5}, with 2–3\% increases for each. This suggests that, while project-level knowledge may not frequently enable models to solve a bug on the first attempt, it increases the likelihood of success with subsequent attempts. This is especially important for agentic APR workflows where fixes are often generated through iterative attempts~\cite{yang2024sweagent, 10.1145/3650212.3680384}. While project context is not a silver bullet, it can boost a model's ability to converge on a correct fix when earlier layers fail.

The impact of project-level knowledge varies across bug types. As shown in Table~\ref{tab:overall_results_bug_type}, Llama sees improvements in 10 additional \textit{Program Anomaly}, 1 \textit{Network}, 3 \textit{GUI}, and 1 \textit{Configuration} bug. GPT-4o-mini, in comparison, fixes 3 more \textit{Program Anomaly}, 1 \textit{Network}, and 3 \textit{GUI} bugs. Neither model shows any improvement on \textit{Performance} or \textit{Permission/Deprecation} bugs in this layer.

%\noindent
\textbf{Discussion.} Project-related knowledge, such as documentation and past resolved issues, offers nuanced, high-level context that can be valuable for certain types of bugs that require understanding of API behavior, developer intent, or project-specific usage constraints. However, its benefits seem more limited for bugs requiring low- to mid-level understanding of the system, such as performance tuning. %or permission/deprecation control, where such resources typically lack detailed coverage. 
Moreover, the effectiveness of this knowledge depends heavily on how well a project maintains its documentation and issue reporting.

\begin{figure}[h]
    \centering
    \includegraphics[width=0.48\textwidth]{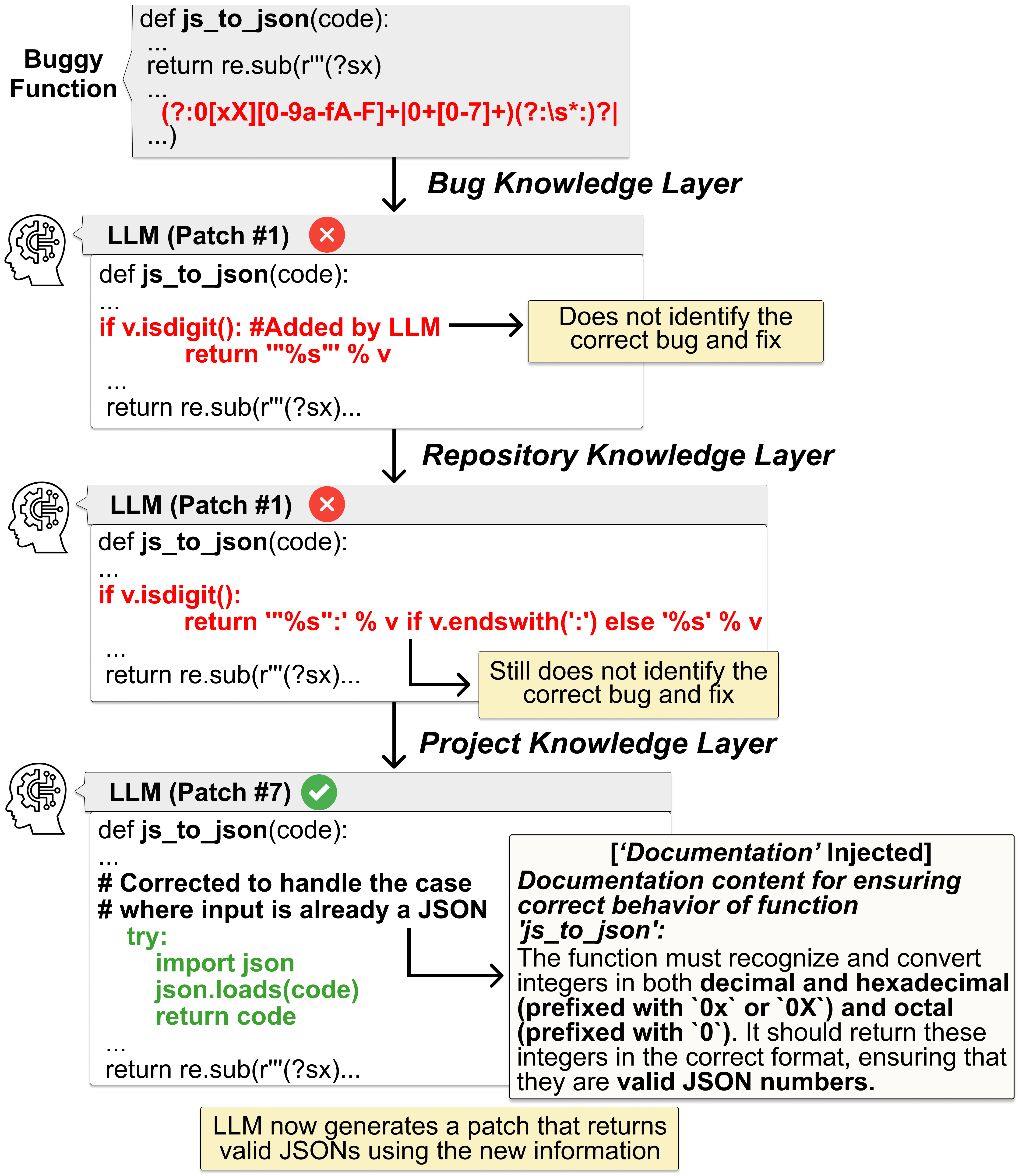}
    \vspace{-0.25cm}
    \caption{Comparison of Patch Generation with Different Layers Injected to LLMs for Bug \#26 in Project Youtube-dl}
    \label{fig:youtube26}
\end{figure}

To better understand the impact of the \textit{Project Knowledge Layer}, we manually analyzed the newly fixed bugs in this layer, 15 for Llama 3.3 and 7 for GPT-4o-mini, and examined the generated patches.
Our analysis followed the same methodology outlined in RQ2, examining how the additional project-level context contributed to successful repair. This was done by the first author, and the findings were refined through collaborative discussions with all authors.
For example, in a \textit{Program Anomaly} in the youtube-dl project~\cite{noauthor_bug_youtube_26} (Figure \ref{fig:youtube26}), the root cause was a regex pattern failing to handle edge cases like integers beginning with zero. Our documentation queries returned a precise constraint: ``\textit{The function must recognize and convert integers in both decimal and hexadecimal (prefixed with 0x or 0X) and octal (prefixed with 0)}". Combined with the buggy function and relevant API context, this likely enabled the model to generate correct patches.

We have noticed similar behaviors for other bugs. 
In a \textit{Network} bug from scrapy~\cite{noauthor_bug_scrapy_21}, the issue involved handling requests to ``\textit{robots.txt}" that triggered a ``\textit{KeyError}". Past resolved issues were likely useful here. One prior issue described a similar error scenario with the recommendation: ``\textit{Implement better error handling to ensure that the function handles cases where netloc might not exist in self.\_parsers}". With this context, the model correctly applied error handling to resolve the bug.
A similar benefit was observed in a \textit{Configuration} bug~\cite{noauthor_bug_thefuck_11}, where the fix required adjusting how command arguments were preserved. %Injected documentation and past developer discussions in similar patched issues helped disambiguate project-specific behavior. For instance, the 
Information from past resolved issue threads emphasizing concerns such as ``Correct Output Handling”, particularly around capturing error messages, and suggestions like ``Preservation of Original Command Arguments" likely guided the model toward a fix. 

These examples show that project-level context fills in semantic gaps left by local and repository layers. When well-curated, this layer enables LLMs to reason at a higher level, aligning fixes not just with code structure but with project-specific behaviors and expectations.

In addition to our hierarchical injection strategy, we compare our approach with an additional baseline where contextual knowledge from all three layers is injected simultaneously, i.e., at the same time, for each bug. As shown in the last two rows of Table~\ref{tab:overall_results}), both LLMs perform worse for all \textit{Pass@k} metrics when provided with all the layers' information at once. In terms of the overall percentage of bugs fixed, Llama 3.3 has a fix rate of 65\%, identical to its performance using only the first layer, but inferior to the hierarchical configurations that use the repository knowledge and then the project knowledge. GPT-4o-mini performs substantially worse: it fixes 14\% fewer bugs compared to its bug-level configuration. These results suggest that providing all information upfront may prevent the model from distinguishing the most relevant context information for a bug, leading to degraded performance. This further highlights the advantage of our hierarchical approach, which allows the model to solve simpler bugs early using minimal context and reserves broader information for more context-hungry bugs that truly require it.

\subsection{RQ4: Which bug types remain challenging for LLMs, even after injecting all layers of contextual knowledge?}
\textbf{Results.}
After applying all layers of contextual knowledge, GPT-4o-mini still fails to fix 85 bugs, consisting of 58 \textit{Program Anomaly}, 9 \textit{Network}, 7 \textit{GUI}, 8 \textit{Configuration}, 3 \textit{Performance}, and 1 \textit{Permission/Deprecation} bug. Llama leaves 64 bugs unresolved, consisting of 45 \textit{Program Anomaly}, 10 \textit{Network}, 5 \textit{GUI}, 1 \textit{Configuration}, 2 \textit{Performance}, and 1 \textit{Permission/Deprecation} bug.
In total, the union of unresolved bugs across both models results in 99 unique cases, with 50 bugs overlapping, indicating a shared set of challenging bugs that neither model could solve. We analyze the union of all unresolved bugs across both models (99 bugs).

First, we report how many of the five additional pieces of context at the repository-level (\textit{Structural Dependencies, Co-occurring Files, Latest Changes}) and project-level (\textit{Documentation, Issue Resolution History}) were available and thus successfully injected for each unresolved bug (Table~\ref{tab:info_injected}). Of the 99 bugs, 39 received all five pieces of information, 40 were missing one, 18 were missing two, and 2 were missing three. Among the missing elements, \textit{Co-occurring Files} and \textit{Structural Dependencies} were the most frequently absent, suggesting that many of these buggy functions are isolated, either not calling or being called by other parts of the codebase. The same pattern is also observed when looking at the intersection of the unresolved bugs between the two models. Out of the 50 overlapping bugs, only 20 bugs have all the information, 21 were missing one, 8 were missing two, and 1 was missing three, with \textit{Structural Dependencies} and \textit{Co-occurring Files} the most frequently absent information.

% \vspace{-0.4cm}
\begin{table}[h]
\caption{Missing Information in Repository and Project Layers for Unresolved Bugs}
% \vspace{-0.2cm}
\label{tab:info_injected}
\resizebox{\columnwidth}{!}{%
\begin{tabular}{c|c|c|c}
\textbf{\begin{tabular}[c]{@{}c@{}}Unresolved Bugs\\ In both Models\end{tabular}} & \textbf{\begin{tabular}[c]{@{}c@{}}Have All\\ Information\end{tabular}} & \textbf{\begin{tabular}[c]{@{}c@{}}Missing\\ 1 Information\end{tabular}} & \textbf{\begin{tabular}[c]{@{}c@{}}Missing\\ \textgreater{}1 Information\end{tabular}} \\ \hline
\textbf{Union (N=99)} & 39 (39\%) & 40 (41\%) & 20 (20\%) \\ \hline
\textbf{Intersection (N=50)} & 20 (40\%) & 21 (42\%) & 9 (18\%) \\
\end{tabular}%
}
% \vspace{-0.3cm}
\end{table}

Table~\ref{tab:complex} summarizes the comparison of unresolved bugs and fixed bugs by complexity metrics. Unresolved bugs consistently exhibit higher complexity across all metrics. %(\textit{Cyclomatic Complexity}, \textit{Lines of Code (LOC)}, \textit{Halstead Volume}, \textit{Halstead Difficulty}, and \textit{Halstead Effort}), 
To quantify the magnitude of difference, we report Cohen’s d. For example, \textit{Cyclomatic Complexity} shows a Cohen’s d of –0.24, indicating a small to medium effect size with higher complexity in unresolved bugs. Among these, the difference in \textit{Lines of Code} is statistically significant (p-value=0.02), while \textit{Cyclomatic Complexity} and \textit{Halstead Difficulty} are borderline significant with p-values of 0.05 and 0.06, respectively. The \textit{Maintainability Index} is higher for fixed bugs, but the difference is not statistically significant based on the test.

%\noindent
\textbf{Discussion.} Unresolved bugs are not randomly distributed; they cluster around specific bug types and higher complexity profiles. In particular, \textit{Program Anomaly}, \textit{Network}, and \textit{GUI} bugs remain the most challenging for both models, even after all layered contextual knowledge is injected. These categories often require a deeper understanding of data flow, or even network and user-facing behavior that is not always available. There is a level of interaction and feedback needed, especially in the case of \textit{GUI} and \textit{Network} bugs, to guide the generation of plausible patches with LLMs. In many cases, the buggy function is also too isolated to connect with useful information elsewhere in the codebase. Over 66\% of unresolved bugs lacked one or more pieces of injected repository or project-level context, limiting the model’s ability to see the full picture. An example of this can be seen in a \textit{GUI} bug from the matplotlib project~\cite{noauthor_bug_matplotlib_2}. The buggy function in this case is 209 lines long, has moderate complexity, and is structurally independent from the rest of the codebase. Fixing a bug at this level likely requires an interactive approach that allows the model to observe the effects of its edits. This showcases the limitations of static prompting and suggests that integrating LLMs into agentic or feedback-driven APR systems, where edits can be tested, observed, and refined, may be essential for handling such bugs.

This observation is further strengthened by our complexity analysis. Unresolved bugs consistently exhibit higher \textit{Cyclomatic Complexity} and \textit{Lines of Code}. This suggests that bugs involving more branching logic, nested control structures, or longer functions may be harder for LLMs to reason about. Higher \textit{Halstead} metrics for unresolved bugs indicate increased cognitive load, reinforcing the idea that LLMs struggle with dense, logic-heavy functions that require precise coordination across multiple code elements.
These results highlight a key limitation of current LLMs: while they can pattern-match and generate plausible fixes in simpler scenarios, they fall short when deeper semantic understanding or multi-step reasoning is required. Addressing these limitations will require advances in model reasoning capabilities and how contextual information is retrieved for isolated and specific types of bugs.

\begin{table}[h]
% \vspace{-0.3cm}
\caption{Complexity of Unresolved Bugs. Statistically Significant (\textbf{*})}
\label{tab:complex}
\resizebox{\columnwidth}{!}{%
\begin{tabular}{c|c|c|c|c}
\textbf{Metric} & \textbf{\begin{tabular}[c]{@{}c@{}}Fixed Bugs\\ Mean\end{tabular}} & \textbf{\begin{tabular}[c]{@{}c@{}}Unresolved Bugs\\ Mean\end{tabular}} & \textbf{Cohen's d} & \textbf{\begin{tabular}[c]{@{}c@{}}u-test\\ P-value\end{tabular}} \\ \hline
\textbf{\begin{tabular}[c]{@{}c@{}}Cyclomatic\\ Complexity\end{tabular}} & 9.29 & 12.14 & \begin{tabular}[c]{@{}c@{}}-0.24\\ (Unresolved\textgreater{}Fixed)\end{tabular} & \textbf{0.05} \\ \hline
\textbf{\begin{tabular}[c]{@{}c@{}}Lines of Code\\ (LOC)\end{tabular}} & 49.8 & 65.32 & \begin{tabular}[c]{@{}c@{}}-0.24\\ (Unresolved\textgreater{}Fixed)\end{tabular} & \textbf{0.02*} \\ \hline
\textbf{\begin{tabular}[c]{@{}c@{}}Halstead\\ Volume\end{tabular}} & 136.4 & 168.34 & \begin{tabular}[c]{@{}c@{}}-0.11\\ (Unresolved\textgreater{}Fixed)\end{tabular} & 0.10 \\ \hline
\textbf{\begin{tabular}[c]{@{}c@{}}Halstead\\ Difficulty\end{tabular}} & 2.63 & 3.08 & \begin{tabular}[c]{@{}c@{}}-0.15\\ (Unresolved\textgreater{}Fixed)\end{tabular} & \textbf{0.06} \\ \hline
\textbf{\begin{tabular}[c]{@{}c@{}}Halstead\\ Effort\end{tabular}} & 954.0 & 1323.63 & \begin{tabular}[c]{@{}c@{}}-0.11\\ (Unresolved\textgreater{}Fixed)\end{tabular} & 0.08 \\ \hline
\textbf{\begin{tabular}[c]{@{}c@{}}Maintainability\\ Index\end{tabular}} & 73.4 & 72.57 & \begin{tabular}[c]{@{}c@{}}0.05\\ (Unresolved\textless{}Fixed)\end{tabular} & 0.73
\end{tabular}%
}
\vspace{-0.2cm}
\end{table}

%% file: 05_threats.tex
\section{Threats to Validity}
\textbf{Construct Validity.}
A threat to construct validity might arise
from the manual bug type annotations. To mitigate this threat, we used an iterative coding process including multiple rounds of discussion and conflict resolution. Each annotator has 3+ years of experience in programming and qualitative research. We also observed a high Cohen’s Kappa inter-rater agreement of 0.8, indicating strong agreement~\cite{keppaArticle}.

%\noindent
\textbf{Internal Validity.}
The training data of proprietary LLMs is not fully disclosed. This limits our ability to verify whether specific bugs or repository content existed in the models' pre-training data. However, our study focuses on the relative effectiveness of different knowledge injection layers, not on absolute performance. For example, for a given bug, if the repository knowledge layer leads to a successful fix where the bug knowledge layer does not, the observed improvement is attributed to the additional injected context, regardless of whether the model has seen the bug before. This design isolates the impact of contextual knowledge and mitigates the influence of potential leakage.
To further strengthen our analysis, we evaluate two architecturally distinct models (Llama 3.3 and GPT-4o-mini), reducing the likelihood that any results hinge on the behavior of a single LLM. Moreover, a recent investigation into benchmark leakage~\cite{zhou2025lessleakbenchinvestigationdataleakage} using the LLM \textit{StarCoder} found only 11\% leakage rate in the BugsInPy dataset. While our study does not use \textit{StarCoder}, it is likely that if leakage exists, it is limited and unlikely to dominate outcomes.
Lastly, while we curated our layered knowledge based on insights from the APR literature, our layers do not represent an exhaustive set of all potentially useful information. Future work may explore alternative or additional forms of contextual input to further refine LLM-based APR.

%\noindent
\textbf{External Validity.}
Our findings may not generalize to all LLMs. However, to improve generalizability, we evaluated two distinct models: Llama 3.3 (70B parameters) and GPT-4o-mini ($\sim$8B parameters). This allows us to test our framework across models with different scales and capabilities~\cite{Banerjee23}, and observe consistent trends in performance improvements across knowledge layers.
We also acknowledge that the distribution of bug types in our dataset is imbalanced. Underrepresented categories (e.g., \textit{Permission/Deprecation}) may not yield insights that are broadly applicable to those bug types. %This limitation highlights the need for future benchmarks that provide more balanced and comprehensive coverage across bug types to support stronger generalization.

%% file: 06_background.tex
\section{Related Work}
\textbf{Prompt Engineering and In-context Learning for APR.}
Recent research has explored a wide range of techniques to enhance the effectiveness of LLMs in fixing software bugs, including prompt engineering, fine-tuning, and context injection. Xia et al.~\cite{10.1109/ICSE48619.2023.00129} first showed that prompting LLMs with buggy functions, without any additional context or customization, outperforms traditional APR tools.
% Subsequent works built on this insight, showing that both fine-tuning and careful prompt design can further improve repair accuracy.
Building on this, MMAPR\cite{10.1145/3649850} and RING~\cite{10.1609/aaai.v37i4.25642} applied prompt learning for syntactic and multi-language repair, while InferFix~\cite{10.1145/3611643.3613892} used few-shot prompting to fix static analysis issues. These early systems focused on local bug context and overlooked broader project-level information.
The limitations of LLMs became more apparent with the release of the SWE-Bench dataset~\cite{SWE-bench2024}, which includes real-world bugs from large projects. Initial attempts showed low success rates (around 2\%), even when LLMs were provided with the localized buggy functions and asked to generate repairs~\cite{SWE-bench2024}. %, revealing that current models struggle to recover the deeper context needed for effective repair.
Ehsani et al.~\cite{ehsani2025detectingpromptknowledgegaps, ehsani2025makeschatgpteffectivesoftware} found that different prompting strategies (e.g., chain-of-thought, tree-of-thought) do not yield substantial improvements in bug resolution tasks, suggesting that merely tweaking prompt structure is not enough.
To address this, researchers explored integrating richer context.
Fan et al.~\cite{10.1109/ICSE48619.2023.00128} showed that fault localization signals improve fix accuracy.
ChatRepair~\cite{10.1145/3650212.3680323} introduced interactive prompting using failing test names and assertions.
Other studies found that including local bug context~\cite{10.1145/3597503.3639086}, relevant identifiers~\cite{10.1109/ICSE48619.2023.00125,10.1109/ASE56229.2023.00047}, and structured cues from stack traces~\cite{52980, haque-etal-2025-towards} improves repair quality.
Contextual cues from bug reports and issue descriptions have also been shown to enhance performance~\cite{10.1145/3639478.3643526}, as has incorporating historical signals such as previous blame commits~\cite{shi2025hafixhistoryaugmentedlargelanguage}.
Building on these ideas, Parasaram et al.~\cite{parasaram2024factselectionproblemllmbased} proposed MANIPLE, a tool that selects and adds bug-related facts into LLM prompts. Their results showed significant gains over naive prompting, especially when space in the context window was limited.

\textbf{Agentic and Iterative Workflows for APR.}
%Beyond static prompting, recent work has explored agent-based~\cite{yang2024sweagent} and iterative workflows to push the boundaries of LLM-based APR. These systems use multiple rounds of querying, reasoning, and context retrieval to navigate large codebases and refine patches over time~\cite{10.1145/3715004}.
Recent work has moved beyond static prompting toward agent-based~\cite{yang2024sweagent} and iterative workflows to enhance LLM-based APR. These systems perform multi-round querying, reasoning, and context retrieval to explore large codebases and refine patches~\cite{10.1145/3715004}. OpenHands~\cite{wang2025openhandsopenplatformai} combines multi-query prompting with agents in sandbox environments to test and validate fixes. AutoCodeRover~\cite{10.1145/3650212.3680384} uses hierarchical code search to support bug localization and patch generation. Both tools show performance gains on SWE-Bench, highlighting the value of interactive, repository-aware workflows.
%OpenHands~\cite{wang2025openhandsopenplatformai} uses multi-query prompting to extract relevant information from the codebase and employs agents interacting with sandbox environments to test hypotheses and validate repairs. Similarly, AutoCodeRover~\cite{10.1145/3650212.3680384} integrates hierarchical code search to retrieve relevant context for bug localization and patch generation. Both tools improve performance on the SWE-Bench benchmark, showcasing the value of interactive, repository-aware workflows.
Iterative approaches have also proven promising. Ruiz et al.~\cite{ruiz2025artrepairoptimizingiterative} used self-iterative prompting, allowing LLMs to refine patches across up to 10 rounds, yielding up to 78\% improvement in generating plausible patches. 
%where LLMs refine their patches across multiple rounds. By limiting each bug to a maximum of 10 attempts, the system achieved up to 78\% improvement in generating plausible patches. 
Similarly, Yang et al. enhanced prompt quality and precision by integrating repository-level knowledge graphs~\cite{yang2025enhancingrepositorylevelsoftwarerepair}. 
%Yang et al.~\cite{yang2025enhancingrepositorylevelsoftwarerepair} proposed using knowledge graphs extracted from the repository to enhance prompt relevance and repair precision.
Other works have explored integrating multiple LLMs and tool agents into a unified APR pipeline~\cite{zhao2024enhancingautomatedprogramrepair}. 
% For example, Zhao et al.~\cite{zhao2024enhancingautomatedprogramrepair} built a framework combining agentic planning with issue logs and extracting solution design, resulting in improved bug localization and fix generation.

%Overall, while these works advanced LLM-based APR, they often (1) limit context to the immediate bug without integrating broader repository or project-level knowledge, (2) treat all bugs uniformly without addressing their varying contextual needs, and (3) rely on black-box agentic methods that lack interpretability. We address these gaps through a transparent, deterministic, and layered knowledge injection framework that incrementally adds bug, repository, and project-level context, while enabling fine-grained analysis of how different bug types benefit from specific contextual information.
Overall, while prior work advanced LLM-based APR, it often (1) limits context to the immediate bug, (2) treats all bugs the same, and (3) relies on black-box agentic methods that lack interpretability. We address these gaps with a transparent, deterministic, layered knowledge injection framework that incrementally adds bug, repository, and project-level context, enabling fine-grained analysis across bug types.

%Overall, while these works have significantly advanced LLM-based APR, they share key limitations: (1) Most of them focus \textit{on immediate bug context, without precisely integrating broader repository and project-level knowledge}, such as structural dependencies and documentation that could improve repair quality. (2) They also treat all bugs the same, \textit{without accounting for the different contextual needs for each bug type}. For instance, our study found that XX information is very effective in solving \textit{Program Anomaly} bugs. (3) Recent agentic approaches, while effective, often operate as \textit{black boxes, which makes them non-deterministic, hard to interpret, and difficult to control}. In this paper, we address these gaps through a layered knowledge injection framework that incrementally provides LLMs with bug, repository, and project-level context. Our method is deterministic, transparent, and fully interpretable, offering fine-grained control over what knowledge is added and when. Furthermore, we analyze performance across different bug types, revealing how distinct bug categories benefit from different forms of contextual information.

%% file: 07_conclusions.tex
\section{Conclusion and Future Work}
We introduced a layered knowledge injection framework for LLM-based APR, addressing key limitations in previous work that focused narrowly on isolated local bug context while ignoring the broader structure of real-world software systems. Our approach iteratively injects \textit{bug}, \textit{repository}, and \textit{project}-level contextual information into prompts, enabling more effective and context-aware repair across a diverse set of bug types.
Our results confirm the hypothesis posed in our motivating example: one of the missing pieces in LLM-based APR is a deeper understanding of repository and project context.
We observed consistent gains in both \textit{\#fixed} and \textit{Pass@k} scores across layers for both Llama 3.3 and GPT-4o-mini, with Llama achieving a 79\% fix rate, a significant improvement of 23\% over prior work~\cite{parasaram2024factselectionproblemllmbased}. All bug types showed improvement from \textit{Bug Knowledge Layer} to \textit{Repository Knowledge Layer}, while only a subset (\textit{Program Anomaly}, \textit{Network}, and \textit{GUI}) further benefited from \textit{Project Knowledge Layer}.
Injecting all contextual information at once underperformed our layered strategy for all bug types, reinforcing the importance of incremental, context-aware knowledge injection for maximizing repair performance.

Error analysis shows that the remaining unresolved bugs are more complex, structurally isolated, or dependent on particular behavior like user interactions and edge-case reasoning, especially for \textit{Program Anomaly}, \textit{Network}, and \textit{GUI} bugs. Future work will focus on integrating our approach with agentic workflows, such as OpenHands~\cite{wang2025openhandsopenplatformai}, which can actively navigate codebases and incorporate feedback during repair. Combining structured context injection with dynamic, feedback-driven reasoning could lead to more adaptive and effective repair systems. Additionally, project-specific fine-tuning on relevant repositories can further improve model alignment with code conventions and domain-specific behavior, leading to more accurate and reliable repairs in complex software systems.

% The results across all layers indicate the 
% Llama is more effective at leveraging layered contextual knowledge. Its consistently higher \textit{Pass@k} scores show greater reliability and robustness, particularly in scenarios where multiple attempts may be needed to converge on a correct fix. This consistency, paired with its superior fix rate on different bug types, suggests that it generalizes better when rich, multi-level context is provided.

% Across both models, we observe that Llama achieves the best performance in the \textit{Project Knowledge Layer} for the majority of bug types, including \textit{Program Anomaly}, \textit{Network}, \textit{GUI}, and \textit{Configuration} bugs. For \textit{Performance} bugs, its highest success rate occurs in the \textit{Repository Knowledge Layer}, while \textit{Permission/Deprecation} bugs are best addressed using only local bug context in the \textit{Bug Knowledge Layer}.
% Interestingly, while GPT-4o-mini achieves a slightly higher overall fix rate for \textit{Network} bugs in the final layer, Llama consistently outperforms it in \textit{Pass@k} scores, indicating more reliable patch generation across multiple attempts.